\documentclass[%
preprint,
11pt,
 amsmath,amssymb,
aps,
]{revtex4-2}

\usepackage{graphicx}% Include figure files
\usepackage{epstopdf}
\usepackage{dcolumn}% Align table columns on decimal point
\usepackage{bm}% bold math
\usepackage{xcolor}
\usepackage{upgreek}
\usepackage{multirow}
\usepackage{tabularx}
\usepackage[abs]{overpic}
\usepackage{placeins}
\usepackage{tikz}
\usepackage{subfig}
\usepackage{comment}

\begin{document}

\title{Role of boundary conditions and wall orientation on the transport of settling inertial particles in wall-bounded flows}

\author{Y. Zhang}
\affiliation{Department of Civil and Environmental Engineering, Duke University, Durham, North Carolina 27708, USA}%Lines break automatically or can be forced with \\

\author{D.H. Richter}%
\affiliation{Department of Civil and Environmental Engineering and Earth Sciences, University of Notre Dame, Notre Dame, IN 46556, USA}

\author{A.D. Bragg}
\email{andrew.bragg@duke.edu}
\affiliation{Department of Civil and Environmental Engineering, Duke University, Durham, North Carolina 27708, USA}%Lines break automatically or can be forced with \\

\begin{abstract}

    Gravitational settling affects particle transport in turbulent flows in two ways; explicitly, by introducing a finite settling velocity, and implicitly, by modifying how the particles interact with the flow field. For wall-bounded flows, when the wall is horizontal (gravity perpendicular to the wall) both the explicit and implicit effects of settling impact the particle transport towards the wall, whereas when the wall is vertical (gravity parallel to the wall) only the implicit effect plays a role. Surprisingly, it was recently demonstrated that even when the settling parameter $Sv$ is very small, settling can play a very significant role in controlling the near-wall transport in a horizontal channel. In this paper, we use direct numerical simulations to explore how this finding is affected by the particle boundary conditions and whether it also occurs in vertical channels where only the implicit effect of settling plays a role. We show that the sensitivity of the particle transport to $Sv$ depends upon the particle boundary conditions, with elastic-collisions (that generate a zero mean particle flux) and absorbing-wall conditions (that generate a negative mean particle flux) exhibiting qualitatively and quantitatively different sensitivities to $Sv$. It is found that for vertical channels the impact of settling on the particle transport is in fact negligible if $Sv$ is small, and only becomes significant when $Sv\geq O(1)$. Finally, we examine the physical mechanisms governing the settling velocity of particles in vertical channel flows, and find that low-speed streaks can play a significant role in suppressing the settling velocity of the particles near the walls, in contrast to the core region where the preferential sweeping mechanism leads to an enhancement of their settling velocity.

\end{abstract}

\maketitle

\section{Introduction}

The transport of small, heavy particulate matter in wall-bounded turbulent flows occurs in many environmental multiphase flows, such as volcanic ash transport \cite{Koyaguchi2009,Bursik2021}, dust storms \cite{zurek93,strausberg05} and pollen transport \cite{Viner2010,Timerman2014}. In these flows, the gravitational settling of the particles plays an important role in their transport towards and away from the surface of the Earth. If the particles are modeled as being inertialess, then under the assumption that they are sufficiently small, the particles will settle on average at their Stokes settling velocity. When their inertia is not negligible, as can be the case for particle transport in the lower portion of the atmospheric boundary layer \cite{richter18}, then their settling velocity depends on their inertia through the role of multiple competing mechanisms \cite{bragg21,Bragg2021}. Many open questions remain regarding understanding and modeling the transport of inertial, settling particles in wall-bounded flows. Discrepancies between observations and modeling of dust size distributions in the planetary boundary layer have been reported \cite{Ryder2013,Rosenberg2014,Adebiyi2020,Evan2014}, and it is possible that some of these discrepancies are due to the models ignoring a number of the physical mechanisms that arise due to the combined effects of particle inertia and settling, as discussed in \cite{bragg21}.

The simplest context for understanding how the combined effects of settling and inertia impact particle transport in turbulence is that of the transport of small (compared to the Kolmogorov scale), heavy (compared to the fluid density) particles in homogeneous turbulence, which can occur far from flow boundaries \cite{pope}. Pioneering studies on this problem are those of Maxey \cite{maxey87} and Wang and Maxey \cite{wang93}, which revealed that inertial particles preferentially sample flow regions where the strain-rate dominates the local velocity gradients compared with the rotation-rate. Gravitational settling introduces an asymmetry into how the particles interact with these regions, leading to a preference for particles to sample downward moving, strain dominated regions of the flow, an effect referred to as the preferential sweeping mechanism. This mechanism leads to an enhancement of the mean particle settling velocity, and is often thought to be strongest when the settling parameter $Sv$ and Stokes number $St$ (both based on the Kolmogorov scales) are $O(1)$. Tom and Bragg \cite{tom19} also developed a multiscale version of the preferential sweeping mechanism to explain in detail the role that different sized eddies play in this process, and how the enhanced settling depends on $St, Sv$ and the flow Taylor Reynolds number $Re_\lambda$.

Bragg, Richter and Wang \cite{Bragg2021} showed that in the context of settling in a horizontal channel flow, the near-wall accumulation of particles is still highly sensitive to $Sv$ even when $Sv\ll 1$. Based on an analysis of the near-wall asymptotics of the particle statistics in the regime $Sv\ll1$, they showed that this occurs because near to the wall, the other mass flux contributions arising due to particle inertia (which dominate further away from the wall when $Sv\ll1$) become sub-leading compared with the Stokes settling flux due to the turbulent fluctuations vanishing at the wall. Therefore, the widespread assumption that settling can be ignored when $Sv\ll1$ is not in general valid, and can therefore play a potentially significant role in wall deposition rates. 

In many problems involving wall-bounded particle transport, for example in atmospheric flows over sloped terrain, or for particle transport through an inclined channel, the wall is not perpendicular to gravity. In the limit of a vertical wall, the explicit effect of gravity on the wall-normal particle transport vanishes. Nevertheless, gravity can still strongly influence the wall-normal transport due to its implicit effect on the way the particles interact with the turbulent flow. Another open question is the extent to which the findings in \cite{Bragg2021} depend on the particular choice of particle boundary conditions. For example, that study considered an absorbing wall boundary condition, but how would the sensitivity to $Sv$ change if elastic particle-wall collisions were considered? The purpose of this paper is to explore these open questions.

\section{Theory}

\subsection{Mechanisms governing wall-normal particle transport}

 In this work, we are interested in small (compared with the Kolmogorov length scale), heavy particles with particle-to-fluid density ratio $\rho_p/\rho_f\gg1$, which are subject to linear drag and gravitational forces \cite{maxey83}. When all variables are expressed in non-dimensional wall-units, the equation reads
 \begin{align}
    \ddot{\bm{x}}^p(t)\equiv\dot{\bm{v}}^p(t) = \frac{1}{St}\Big(\bm{u}^p(t)-\bm{v}^p(t)\Big) + \frac{Sv}{St}\bm{e}_g,  \label{parteq} 
 \end{align}
 where $\bm{x}^p(t)$ and $\bm{v}^p(t)$ are the position and velocity of the particle, respectively, $\bm{u}^p(t)$ denotes the fluid velocity at the particle position, and $\bm{e}_g$ denotes the unit vector in the direction of gravity. In the equation, $St\equiv\tau_p/\tau_*$ is the particle Stokes number based on the fluid friction timescale $\tau_*$, and $Sv\equiv\tau_pg/u_*$ is the particle settling number based on the fluid friction velocity $u_*$. 

 The contribution $(Sv/St)\bm{e}_g$ in the equation of motion corresponds to the explicit gravitational effect which acts in the direction $\bm{e}_g$. However, even for particle motion in the plane orthogonal to $\bm{e}_g$, gravitational settling still impacts the particle motion implicitly through $\bm{u}^p(t)$, since $\bm{u}^p(t)$ depends on the three-dimensional trajectory of the particle $\bm{x}^p(t)$  which is itself dependent on gravity. This impact of gravity on $\bm{u}^p(t)$ corresponds to the implicit effect of gravity discussed above. 
 
 An equation for the mean particle momentum in the wall-normal direction (assuming the flow is homogeneous in the other directions) can be derived from \eqref{parteq} using a phase-space PDF analysis \cite{bragg21}:
 \begin{align}
      StD_t\langle w^p(t)\rangle_z =  \langle u^p(t)\rangle_z - \langle w^p(t)\rangle_z - St\Big(\nabla_zS + \frac{S}{\varrho}\nabla_z\varrho\Big) + Sv\bm{e}_g\cdot\bm{e}_z, \label{evleqn}
 \end{align}
 where $z^p(t)$, $w^p(t)$ and $u^p(t)$ are the projections of  $\bm{x}^p(t)$, $\bm{v}^p(t)$ and $\bm{u}^p(t)$ in the $\bm{e}_z$ (wall-normal) direction, respectively, $D_t\equiv\partial_t + \langle w^p(t)\rangle_z\nabla_z$, $\varrho$ is the particle concentration PDF, $\langle\cdot\rangle_z$ denotes an ensemble average conditioned on $z^p(t)=z$, and $S\equiv\langle[w^p(t)-\langle w^p(t)\rangle_z]^2\rangle_z$ is the variance of the wall-normal particle velocity. 

 In \eqref{evleqn}, $-St\nabla_zS$ corresponds to the turbophoretic drift mechanism that drives the particles towards the wall in the near-wall region. The term $-St(S/\varrho)\nabla_z\varrho$ represents an inertial diffusion mechanism that arises due to the diffusive effect of the decoupling between the particle and fluid velocities. The term $\langle u^p(t)\rangle_z$ denotes the mean wall-normal fluid velocity sampled by the particles, and this is non-zero due to the particles preferentially sampling the flow. As discussed in \cite{Bragg2021}, for a horizontal channel flow, in the quasi-homogeneous region of a boundary layer, this term will act to drive the particles towards the wall and is associated with the preferential sweeping mechanism. Close to the wall this term drives the particles away from the wall, due to the diffusive effect of the turbulent fluctuations on the particle transport. Indeed, for a zero mass flux configuration where $\langle w^p(t)\rangle_z=0$, the positivity of $\langle u^p(t)\rangle_z$ is required near the wall to balance the drift towards the wall due to the turbophoretic drift (and gravitational settling for the case of a horizontal channel). A more detailed discussion of each term appearing in \eqref{evleqn} can be found in \cite{Bragg2021}.

 \subsection{Explicit and implicit gravitational effects in the zero mass flux case}
Corresponding to \eqref{evleqn}, the equation governing the particle concentration is \cite{bragg21}
 \begin{align}
      \partial_t\varrho +\nabla_z\Big(\varrho\langle w^p(t)\rangle_z\Big)=0, \label{Mass_baleqn}
 \end{align}
such that for  a statistically stationary system we have $\varrho\langle w^p(t)\rangle_z=\Phi$, where $\Phi$ is the constant mass flux that is determined by the boundary conditions for the particles. When the mass flux is zero $\Phi=0$, \eqref{evleqn} reduces to
 \begin{align}
      0 =  \langle u^p(t)\rangle_z - St\Big(\nabla_zS + \frac{S}{\varrho}\nabla_z\varrho\Big) + Sv\bm{e}_g\cdot\bm{e}_z. \label{baleqn}
 \end{align}
 For a vertical wall, $\bm{e}_g\cdot\bm{e}_z = 0$ and the explicit gravitational settling contribution vanishes in the wall-normal direction. The implicit effect of settling on the remaining terms persists, however.  For a horizontal wall, $\bm{e}_g\cdot\bm{e}_z = -1$ and both the explicit and implict effects of gravitational settling influence the particle transport.

 Rearranging \eqref{baleqn} and integrating over the region $z\leq H-d_p/2$ (where $H$ is the height of the flow, and $d_p$ is the particle diameter) leads to the following formal solution 
 %
 %\begin{align}
 %    \nabla_z\varrho = \frac{1}{S}\Big( -\frac{Sv}{St} + \frac{1}{St}\langle u^p(t)\rangle_z + \nabla_zS \Big)\varrho, \label{oderho}
 %\end{align}
 \begin{align}
    \begin{split}
        \varrho(z) &=\varrho_0\exp\Bigg(-\bm{e}_g\cdot\bm{e}_z \frac{Sv}{St}\int^{H-d_p/2}_{z} \frac{1}{S} \,dq\Bigg)\exp\Bigg(-\frac{1}{St} \int^{H-d_p/2}_{z} \frac{1}{S}\Big[ \langle u^p(t)\rangle_q- St\nabla_qS \Big] \,dq\Bigg)\\
        &\equiv \varrho_0\exp{(\mathcal{A})}\exp{(\mathcal{B})},\label{rhosln}
    \end{split}
 \end{align}
 where $\varrho_0\equiv\varrho(H-d_p/2)$ which must be chosen such that $\int^{H-d_p/2}_{d_p/2}\varrho(z)\,dz=1$, where , and we have introduced $\mathcal{A}(z)\equiv -\bm{e}_g\cdot\bm{e}_z\frac{Sv}{St}\int^{H-d_p/2}_{z} \frac{1}{S} \,dq$ and $\mathcal{B}(z)\equiv -\frac{1}{St}\int^{H-d_p/2}_{z} \frac{1}{S}\Big( {\langle u^p(t)\rangle_q}- St\nabla_qS \Big) \,dq$ for simplicity of discussion. $\mathcal{A}(z)$ is associated with the gravitational settling contribution in \eqref{baleqn} while $\mathcal{B}(z)$ is associated with the difference between the preferential sampling and turbophoretic drift terms, $\langle u^p(t)\rangle_z-St\nabla_zS$. We will now consider the contributions from $\mathcal{A}(z)$ and $\mathcal{B}(z)$ to the behavior of $\varrho$.

 When $Sv=0$, $\mathcal{A}(z)=0$ and Sikovsky \cite{sikovsky14} showed that $\varrho$ exhibits a power-law singularity in the limit $z\to 0$ (except when $St$ is very large, in which case $\varrho$ is constant in this limit), and that this comes from the term involving $\mathcal{B}(z)$. To understand the role of $\mathcal{A}(z)$ we can consider the regime $Sv\ll1$ and suppose that in this regime the effect of settling is perturbative. Then, the quantities $S$ and $\langle u^p(t)\rangle_z$ can be expanded as a regular perturbation series in $Sv$, with the leading-order terms independent of $Sv$. Let us denote these leading-order contributions using the superscript $[0]$. According to the analysis of \cite{sikovsky14,johnson20}, $S^{[0]}\sim a z^{4}$ and $\langle u^p(t)\rangle_z^{[0]}\sim b z^3$ for $St\ll 1$ and $z\ll 1$, where $a,b$ are positive and depend on $St$ but not $z$. Using these, the asymptotic behavior of \eqref{rhosln} for $z\ll 1$ (with $z\ll H$) and for $St\ll 1, Sv\ll1$ is 
 
   \begin{align}
    \begin{split}
        \varrho(z) &\sim \varrho_0\Bigg(\frac{z}{H}\Bigg)^{\frac{b-4a St}{aSt}}\exp\Bigg(- \frac{Sv}{3 a St}\bm{e}_g\cdot\bm{e}_z z^{-3}\Bigg).\label{rhosln_lowStSv}
    \end{split}
 \end{align}
When $St\geq O(1)$ and inertial effects dominate the particle motion (but not so large that the particles move ballistically through the boundary layer), the analysis of \cite{sikovsky14} shows that $S^{[0]}\sim a z^{\gamma}$ with $\gamma(St)\in[3,4]$, and the contribution from $\langle u^p(t)\rangle_q$ in \eqref{rhosln} is sub-leading. In this case, the asymptotic behavior of \eqref{rhosln} for $z\ll 1$ (with $z\ll H$) and for $St\geq O(1), Sv\ll1$ is

   \begin{align}
    \begin{split}
        \varrho(z) &\sim \varrho_0\Bigg(\frac{z}{H}\Bigg)^{-\gamma}\exp\Bigg(- \frac{Sv}{a(1-\gamma) St}\bm{e}_g\cdot\bm{e}_z z^{1-\gamma}\Bigg).\label{rhosln_highStlowSv}
    \end{split}
 \end{align}
The power-law contributions in \eqref{rhosln_lowStSv} and \eqref{rhosln_highStlowSv} correspond to those discussed \cite{sikovsky14,johnson20} for the case of non-settling particles, and describe the near-wall particle accumulation due to the turbophoretic mechanism that causes the particles to drift towards the wall. This contribution is independent of the orientation of the wall in the regime $Sv\ll1$. The exponential contributions in \eqref{rhosln_lowStSv} and \eqref{rhosln_highStlowSv} correspond to the near-wall accumulation driven by gravitational settling when $\bm{e}_g\cdot\bm{e}_z\in [-1,0)$. These contributions only remain finite when the wall is not vertical, i.e. when $\bm{e}_g\cdot\bm{e}_z\neq 0$. Since these contributions grow as exponentials of power-law functions of $z$, they increase rapidly in the limit $z\to 0$ and quickly dominate over the power-law contribution to $\varrho(z)$ in this limit. Hence, even if $Sv\ll1$, there will always exist a region sufficiently close to the wall where the gravitational contribution to $\varrho(z)$ dominates over the power-law contribution due to inertia, and hence in this region, settling is not in fact a perturbative contribution to the particle motion (contrary to the naive assumption that this follows due to $Sv\ll1$). This is the essence of the argument put forth in \cite{Bragg2021} for the particular case of a horizontal wall where  $\bm{e}_g\cdot\bm{e}_z=-1$.

It is important to note that although the analysis above only corresponds to the near wall region $z\ll1$, it also indirectly implies something about the behavior of $\varrho(z)$ further away from the wall. This is because as a PDF it must satisfy the integral constraint $\int^{H-d_p/2}_{d_p/2}\varrho(z)\,dz=1$. Hence, if settling strongly impacts $\varrho(z)$ near the wall even when $Sv\ll 1$, it must also affect it further away from the wall in order to preserve $\int^{H-d_p/2}_{d_p/2}\varrho(z)\,dz=1$.

\subsection{Explicit and implicit gravitational effects in the finite mass flux case}

For the case where the mass flux $\Phi$ is finite, e.g. due to an absorbing wall boundary condition, the formal solution  to \eqref{evleqn} is 
 \begin{align}
     \varrho(z)= \varrho_0\exp{(\mathcal{A})}\exp{(\mathcal{B})}\Bigg(1 + \frac{\Phi}{St}\int_z^{H-d_p/2} \frac{1}{S}\exp[-\mathcal{A}(q)-\mathcal{B}(q)]\, dq\Bigg). \label{rhosln_elas}
 \end{align}
 When $\Phi=0$, this result reduces to \eqref{rhosln}, but in general it contains an additional contribution due to $\Phi$. The value of $\Phi$ and its dependence on the system parameters is determined by the boundary conditions for the particles. One case of interest is particle transport in a high Reynolds number turbulent boundary layer (e.g. the atmospheric boundary layer) where $H$ is so large that at $z=O(H)$ the effect of particle inertia negligible and $\langle w^p(t)\rangle_{H-d_p/2}=  Sv\bm{e}_g\cdot\bm{e}_z$, leading to 

 \begin{align}
     \varrho(z)= \varrho_0\exp{(\mathcal{A})}\exp{(\mathcal{B})}\Bigg(1 + \varrho_0\frac{ Sv}{St}\bm{e}_g\cdot\bm{e}_z\int_z^{H-d_p/2} \frac{1}{S}\exp[-\mathcal{A}(q)-\mathcal{B}(q)]\, dq\Bigg).\label{rhosln_absp}
 \end{align}
In this case we clearly see that the presence of a finite mass flux due to settling introduces an additional dependence upon $Sv$ into the solution for $\varrho(z)$ compared with the zero-flux case. For this reason, the sensitivity of $\varrho(z)$ to $Sv$ for a vertical channel may differ from that for a horizontal channel (the case which was the case explored in \cite{Bragg2021}). Since $\frac{1}{S}\exp[-\mathcal{A}(q)-\mathcal{B}(q)]\geq 0$ then the contribution $\varrho_0\frac{ Sv}{St}\bm{e}_g\cdot\bm{e}_z\int_z^{H-d_p/2} \frac{1}{S}\exp[-\mathcal{A}(q)-\mathcal{B}(q)]\, dq$ will be negative when $\bm{e}_g\cdot\bm{e}_z\in [-1,0)$. Hence the contribution to $ \varrho(z)$ arising from the finite flux $\Phi$ will act to reduce the near-wall accumulation of the particles compared to the zero-flux case when $\bm{e}_g\cdot\bm{e}_z\in [-1,0)$. The impact of settling on $\varrho(z)$ may therefore be very different depending on $\Phi$, e.g. due to the additional contribution of settling to $\varrho(z)$ through $\varrho_0\frac{ Sv}{St}\bm{e}_g\cdot\bm{e}_z\int_z^{H-d_p/2} \frac{1}{S}\exp[-\mathcal{A}(q)-\mathcal{B}(q)]\, dq$ in the case of a high Reynolds number boundary layer. Intuitively, this makes sense because for elastic collisions that generate $\Phi=0$, sufficiently strong settling may cause the particles to become trapped near the wall, since only very strong turbulent fluctuations will be able to resuspend them into the flow. Hence, settling will lead to a strong accumulation of the particles at the wall, stronger than for the  $Sv=0$ case. On the other hand, for an absorbing horizontal wall that generates $\Phi < 0$ in the presence of settling, strong settling will simply cause the particles to rapidly fall through the viscous sublayer before being absorbed by the wall. In this case settling suppresses the ability of the particles to linger in the viscous sublayer where the turbulent fluctuations are weak, which is precisely what allows for their strong accumulation in the $Sv=0$ case.

\section{Direct Numerical Simulations}

To investigate the influence of boundary conditions and the explicit and implicit effects of settling, we used DNS to track the motion of settling, inertial particles in an incompressible closed channel flow. For the fluid phase, the momentum equation together with the continuity equation were solved
\begin{align}
    \partial_t\bm{u} + (\bm{u\cdot\nabla})\bm{u} &= -\frac{1}{\rho_f}\bm{\nabla}p + \nu\nabla^2\bm{u},\\
    \bm{\nabla\cdot u} &= 0,
\end{align}
where $\bm{u}$ is the fluid velocity, $p$ is pressure, and $\nu$ is the kinematic viscosity of the fluid. The fully developed channel flow is driven by a constant mean pressure gradient and is periodic in the streamwise and spanwise directions, with the spatial coordinates in these directions denoted by $x$ and $y$, respectively. The wall-normal spatial coordinate is $z$, and at $z=0$ (lower wall) and $z=L_z=2H_{1/2}$ (upper wall, where $H_{1/2}$ is the half channel height), no-slip conditions are imposed for the velocity field. The domain has a size of $L_x\times L_y\times L_z = 2\pi\times\pi\times1$. A pseudospectral method is used in the periodic directions while a 2nd-order finite difference method is used in the wall-normal direction. Time stepping of the fluid velocity and pressure fields is achieved using a 3rd-order Runge-Kutta method. The channel flows simulated all have a friction Reynolds number of $Re_{\tau}\equiv u_{\tau}H_{1/2}/\nu=315$. 

Lagrangian point particles are tracked in the flow by solving equation \eqref{parteq} using a backward Euler method. The fluid velocity at the particle positions is computed with a sixth-order Lagrangian interpolation method. The particle loading is assumed to be small enough to neglect two-way coupling and particle-particle collisions for consistency with the previous study \cite{bragg21} that this work builds on and compares with.  

{\begin{figure}
    \centering
		{
        \begin{overpic}
            [trim = 0mm 0mm 0mm 0mm,scale=0.58,clip,tics=20]{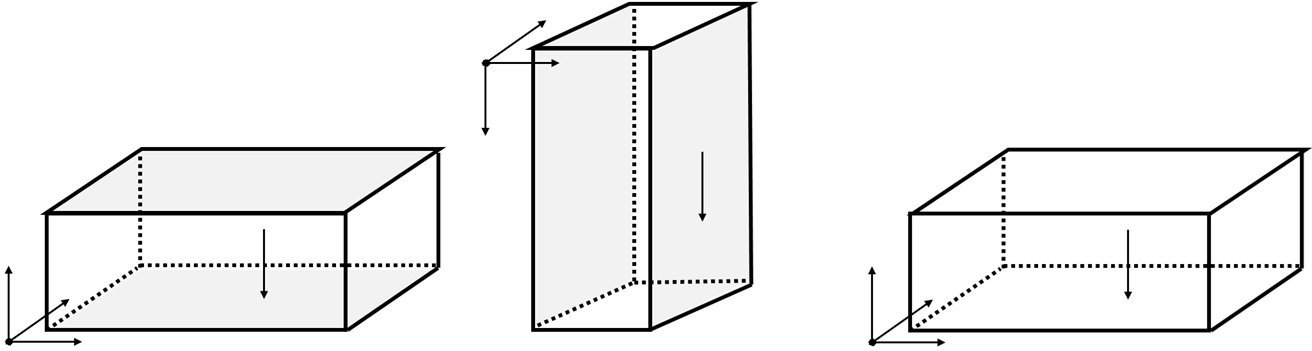}
            \put(10,125){\textbf{(a) Horizontal Channel}}
            \put(55,110){($\Phi=0$)}
            \put(25,0){\small{$x$}}
            \put(-6,23){\small{$z$}}
            \put(18,23){\small{$y$}}
            \put(60,30){\small{$g$}}
            \put(45,63){no-slip wall}
            \put(45,-5){no-slip wall}
            \put(140,125){\textbf{(b) Vertical Channel}}
            \put(180,110){($\Phi=0$)}
            \put(144,60){\small{$x$}}
            \put(165,75){\small{$z$}}
            \put(165,99){\small{$y$}}
            \put(210,40){\small{$g$}}
            \put(220,20){\rotatebox{90}{no-slip wall}}
            \put(160,20){\rotatebox{90}{no-slip wall}}
            \put(270,125){\textbf{(c) Horizontal Channel}}
            \put(315,110){($\Phi<0$)}
            \put(285,0){\small{$x$}}
            \put(254,23){\small{$z$}}
            \put(278,23){\small{$y$}}
            \put(320,30){\small{$g$}}
            \put(305,63){no-slip wall}
            \put(305,-5){no-slip wall}
		\end{overpic}}	
        \caption{The three configurations (a) a horizontal channel with $\Phi=0$ (HZF case), (b) a vertical channel with $\Phi=0$ (VZF case), and (c) a horizontal channel with $\Phi<0$ (HNF case). In these schematics, the shaded planes denote the surfaces where the particles rebound elastically, while the non-shaded planes denote the surfaces where either periodic or absorbing wall boundary conditions apply.} 
        \label{configs}
\end{figure}
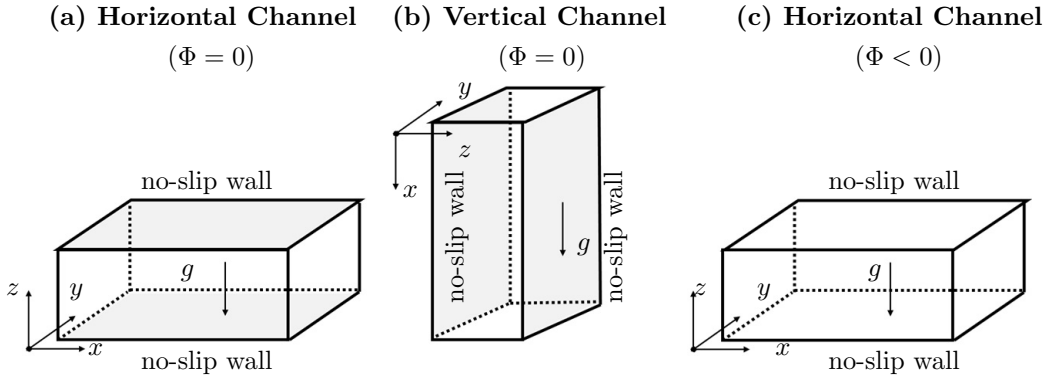}

Three closed channel flow configurations are considered: a horizontal channel with reflecting walls for the particles (i.e. elastic particle-wall collisions), a vertical channel with reflecting side walls for the particles, and a horizontal channel with absorbing upper and lower walls for the particles (see Fig \ref{configs}). In each case, periodic boundary conditions are applied to the particles in the streamwise and spanwise directions. 

For all three configurations, the particles are initially released with a uniform distribution and a velocity equal to their Stokes settling velocity. In the horizontal and vertical channel flows with two reflecting walls (see Fig \ref{configs} (a) \& (b)), the particles remain in the domain throughout the entirety of the simulation. At steady state, these cases generate a zero particle mean mass flux $\Phi=0$. For the horizontal channel with absorbing walls (see Fig \ref{configs} (c)), new particles must be introduced to replace those that are absorbed by the walls in order to maintain a steady state. These new particles are released randomly at the centerline of the domain with a velocity equal to their Stokes settling velocity plus the local fluid velocity. At steady state, the horizontal channel with absorbing walls generates a negative particle mean mass flux $\Phi<0$ due to settling. Hereafter, the horizontal and vertical channel cases with $\Phi=0$ will be referred to by HZF and VZF, respectively, while the horizontal channel cases with $\Phi<0$ will be referred to by HNF.

Note that in \cite{bragg21,Bragg2021} the case with an absorbing wall for the particles used an open channel with an elastically rebounding boundary condition for the particles at the top of the domain. However, for consistency with the other cases explored in this work, a closed channel is used here for the cases with absorbing walls (i.e. the HNF cases). Nevertheless, as we will see, the results for the particle statistics near the lower wall (the region we focus on) using the closed channel are very similar to those in \cite{bragg21,Bragg2021} for an open channel.

In the DNS, four values of $St$ corresponding to weak to intermediate particle inertia were considered $St\in[0.93, 46.5]$, consistent with the values considered in \cite{bragg21,Bragg2021}. For reasons that will be explained in the next section, different values of $Sv$ in the range $Sv\in[0,1]$ were explored for the different cases, with smaller values considered for the HZF cases. The particle statistics are computed using a stretched grid with refined near-wall resolution, with grid points $N_x\times N_y\times N_z = 128\times128\times180$ corresponding to spacing (in wall-units) $\Delta_x^+ \times \Delta_y^+ \times \Delta_z^+ = 30.95\times15.48\times0.25\text{ (wall), } 12.34\text{ (center)}$.

\section{Results and Discussions}

 \subsection{Statistically stationary regimes}

The near-wall particle concentration in a turbulent channel flow can take a significant amount of time to reach a statistically stationary regime. Since the impact of settling when $Sv\ll1$ is most important in the near-wall region, we therefore first check that the results do indeed correspond to a statistically stationary regime. 

In Figure \ref{stat}, two HZF cases [(a) \& (b)] and two HNF cases [(c) \& (d)] are shown. For all $St$ and $Sv$ numbers tested, the maximum particle concentration $\varrho_{max}=\max[\varrho]$ is used to test for statistical stationarity. Since the particles accumulate most strongly near the lower wall in the horizontal channel configurations, at steady state $\varrho_{max}$ is associated with $\varrho$ near or at the lowest grid point. As the plots illustrate, the DNS is run for very long times to ensure that $\varrho_{max}$ reaches an approximate steady state. Once $\varrho_{max}$ has reached its approximate steady state value (which generally occurs by $t\sim O(10^5\tau_*)$), we continue to run the DNS for at least another $1\times10^5\tau_*$ before the particle statistics are constructed by averaging over many snapshots of the particle positions and velocities in this steady state. The second-order particle statistics $\langle[w^p(t)]^2\rangle_z$ and $\langle[u^p(t)]^2\rangle_z$ over this same time range are also in a steady state.

 \subsection{Influence of boundary conditions on concentrations in a horizontal channel}
{\vspace{0mm}\begin{figure}
		\centering
		\subfloat[]
		{\begin{overpic}
				[trim = -10mm -3mm 0mm 0mm,scale=0.42,clip,tics=20]{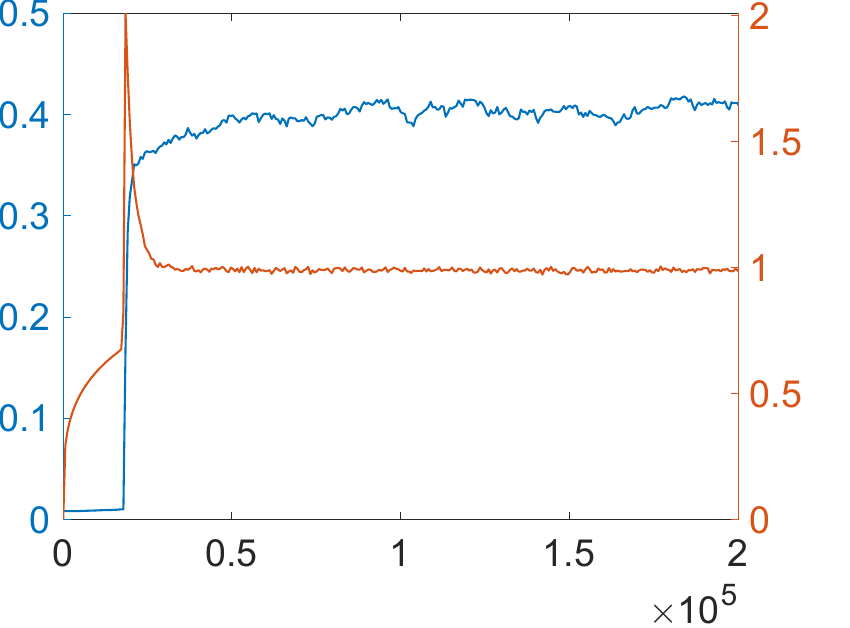}
                
                \put(-5,70){\rotatebox{90}{$\varrho_{max}$}}
                \put(170,70){\rotatebox{90}{$u_\tau$}}
                \put(90,0){$t/\tau_*$}

                \put(95,95){$\varrho_{max}$}
                \put(95,65){$u_\tau$}
									
		\end{overpic}}
		\subfloat[]
		{\begin{overpic}
				[trim = -14mm -3mm -8mm 0mm,scale=0.42,clip,tics=20]{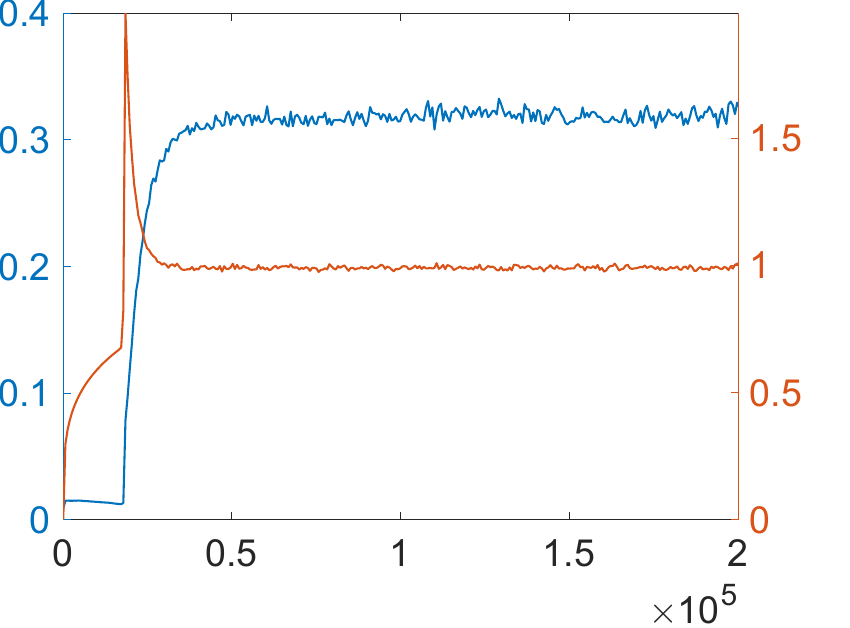}
                
                \put(5,70){\rotatebox{90}{$\varrho_{max}$}}
                \put(175,70){\rotatebox{90}{$u_\tau$}}
                \put(90,0){$t/\tau_*$}

                \put(95,95){$\varrho_{max}$}
                \put(95,65){$u_\tau$}
                
		\end{overpic}}
        \\        
		\subfloat[]
		{\begin{overpic}
				[trim = 0mm -3mm -8mm -3mm,scale=0.42,clip,tics=20]{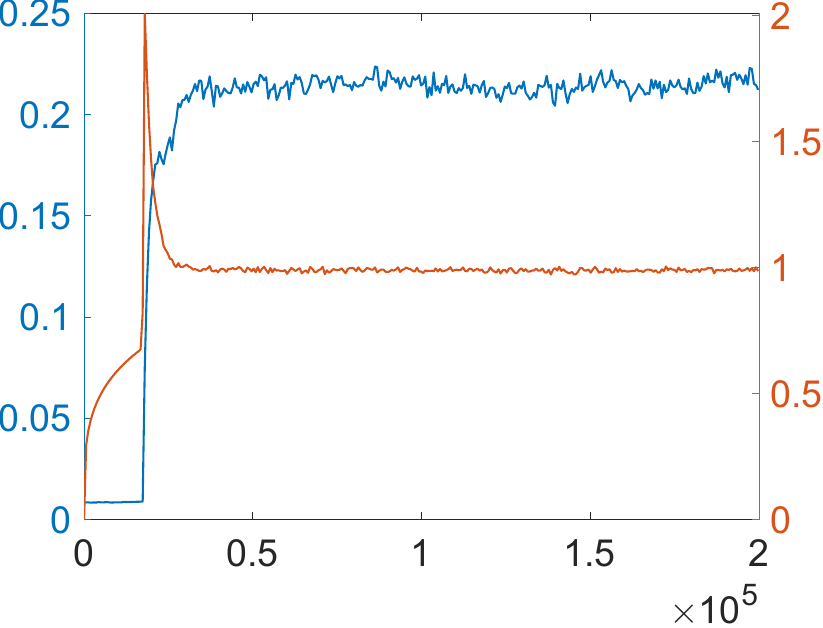}
                
                \put(-15,70){\rotatebox{90}{$\varrho_{max}$}}
                \put(165,70){\rotatebox{90}{$u_\tau$}}
                \put(80,0){$t/\tau_*$}

                \put(95,95){$\varrho_{max}$}
                \put(95,65){$u_\tau$}
                
		\end{overpic}}
		\subfloat[]
		{\begin{overpic}
				[trim = -11mm -5mm 0mm 0mm,scale=0.42,clip,tics=20]{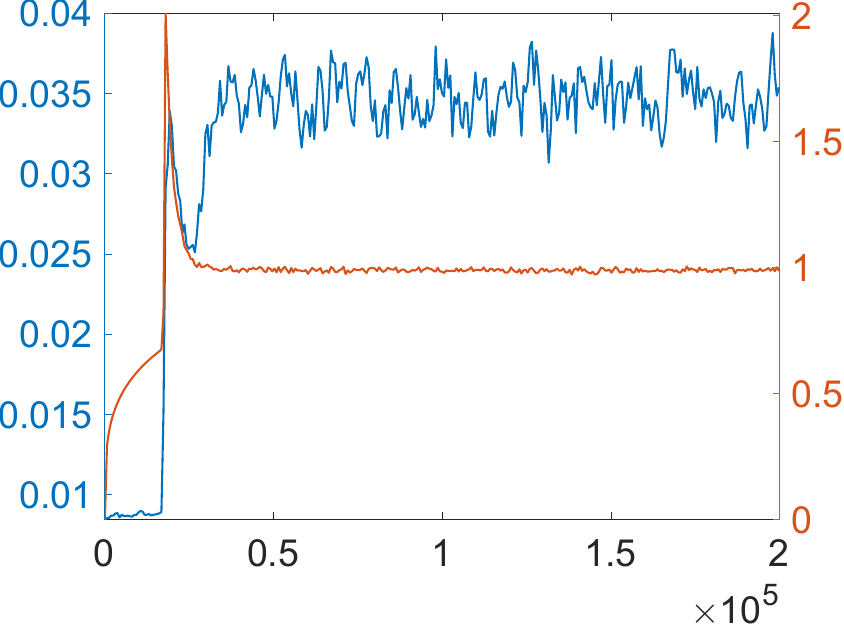}
                
               \put(5,70){\rotatebox{90}{$\varrho_{max}$}}
                \put(180,70){\rotatebox{90}{$u_\tau$}}
                \put(95,0){$t/\tau_*$}

                \put(95,95){$\varrho_{max}$}
                \put(95,65){$u_\tau$}
                
		\end{overpic}}
        \caption{Fluid and particle statistics during the simulation (red curve corresponds to the friction velocity $u_{\tau}$ as a function of time, shown for reference). (a) $\varrho_{max}=\max[\varrho]$ at $St = 4.65$, $Sv = 10^{-5}$ in HZF case, (b) $\varrho_{max}$ at $St = 46.5$, $Sv = 10^{-5}$ in HZF case, (c) $\varrho_{max}$ at $St = 4.65$, $Sv = 10^{-4}$ in HNF case, (d) $\varrho_{max}$ at $St = 46.5$, $Sv = 10^{-4}$ in HNF case.} 
		\label{stat}
\end{figure}}
 {\vspace{0mm}\begin{figure}
		\centering
		\subfloat[]
		{\begin{overpic}
				[trim = 0mm 0mm -15mm 0mm,scale=0.42,clip,tics=20]{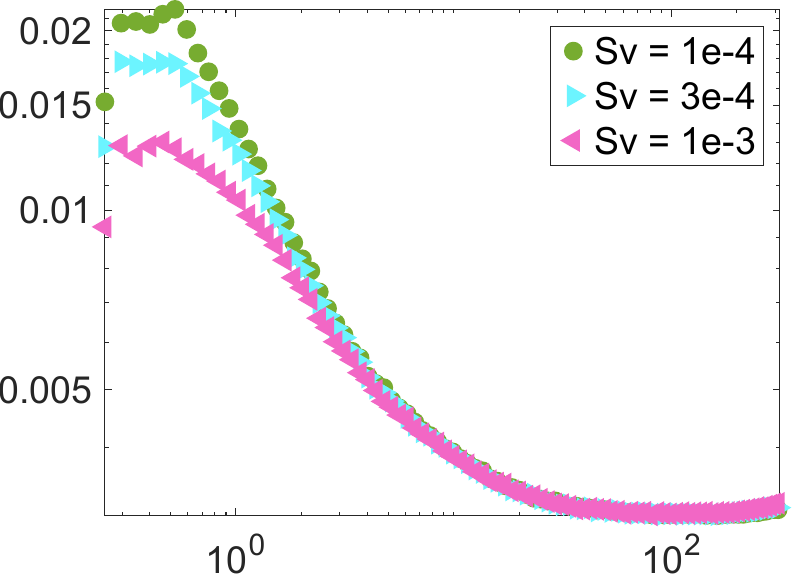}

                %\put(25,140){\textbf{\large{Elastic BC}}}
                
                \put(90,60){$St = 0.93$}
                \put(-10,60){\rotatebox{90}{$\varrho$}}
                \put(90,-5){$z$}
									
		\end{overpic}}
		\subfloat[]
		{\begin{overpic}
				[trim = -15mm 0mm 0mm 0mm,scale=0.42,clip,tics=20]{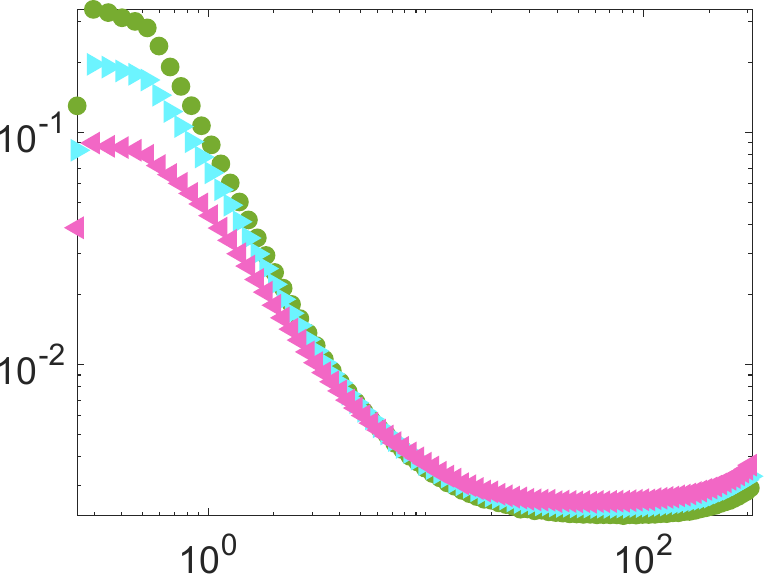}
                
                %\put(35,140){\textbf{\large{Absorbing BC}}}
                
                \put(110,60){$St = 4.65$}
                \put(10,60){\rotatebox{90}{$\varrho$}}
                \put(90,-5){$z$}
                
		\end{overpic}}
        \\
        \subfloat[]
		{\begin{overpic}
				[trim = 0mm 0mm -15mm 0mm,scale=0.42,clip,tics=20]{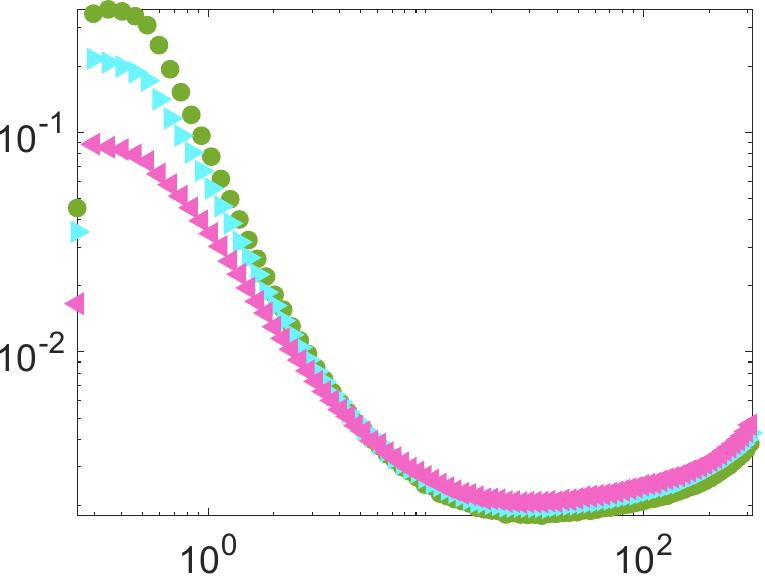}
                
                \put(90,60){$St = 9.3$}
                \put(-10,60){\rotatebox{90}{$\varrho$}}
                \put(90,-5){$z$}
		\end{overpic}}
        \subfloat[]
		{\begin{overpic}
				[trim = -15mm 0mm 0mm 0mm,scale=0.42,clip,tics=20]{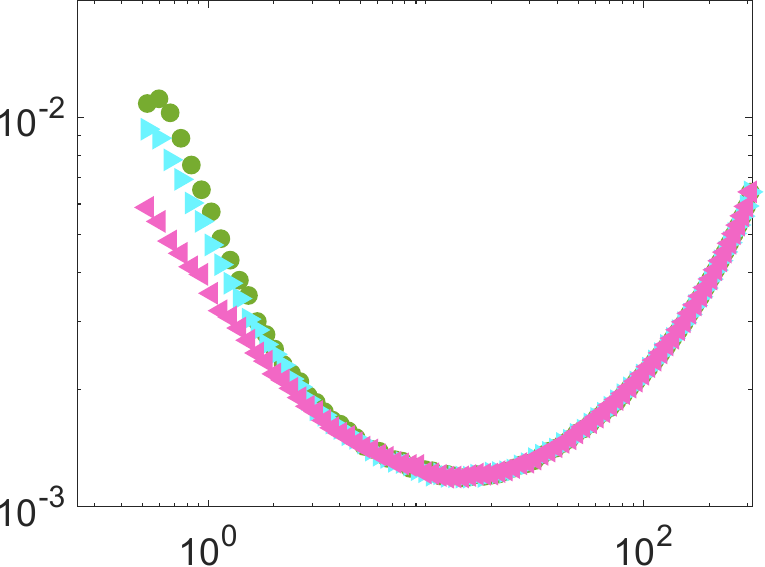}
                
                \put(110,60){$St = 46.5$}
                \put(10,60){\rotatebox{90}{$\varrho$}}
                \put(90,-5){$z$}
		\end{overpic}}
        \caption{Particle concentration $\varrho$ at different $Sv$ for (a) $St=0.93$, (b) $St=4.65$, (c) $St=9.3$, (d) $St=46.5$ in the HNF case.}
		\label{BC_Absp}
\end{figure}}
 {\vspace{0mm}\begin{figure}
		\centering
		\subfloat[]
		{\begin{overpic}
				[trim = 0mm 0mm -15mm 0mm,scale=0.42,clip,tics=20]{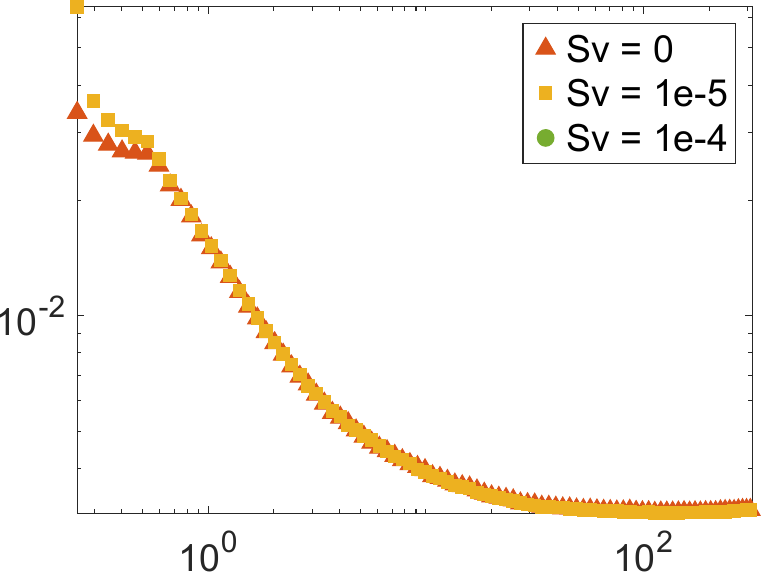}

                %\put(25,140){\textbf{\large{Elastic BC}}}
                
                \put(90,60){$St = 0.93$}
                \put(-10,60){\rotatebox{90}{$\varrho$}}
                \put(90,-5){$z$}
									
		\end{overpic}}
		\subfloat[]
		{\begin{overpic}
				[trim = -15mm 0mm 0mm 0mm,scale=0.42,clip,tics=20]{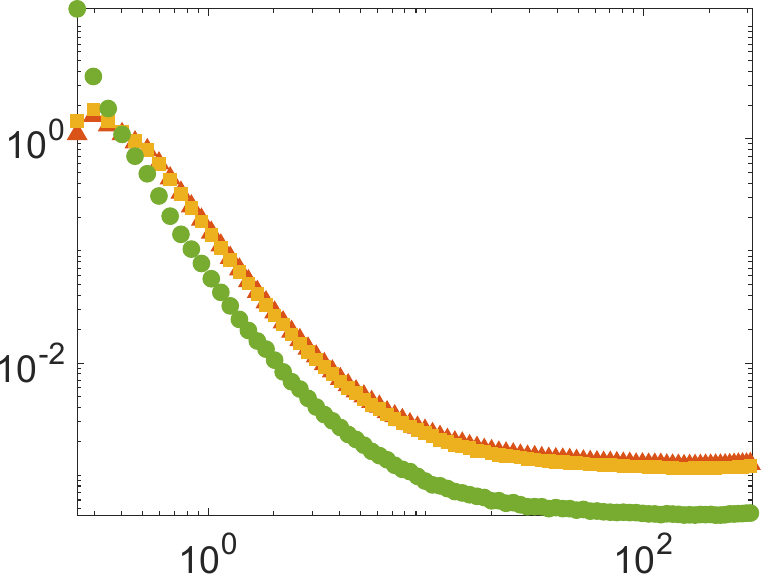}
                
                %\put(35,140){\textbf{\large{Absorbing BC}}}
                
                \put(110,60){$St = 4.65$}
                \put(10,60){\rotatebox{90}{$\varrho$}}
                \put(90,-5){$z$}
                
		\end{overpic}}
        \\
        \subfloat[]
		{\begin{overpic}
				[trim = 0mm 0mm -15mm 0mm,scale=0.42,clip,tics=20]{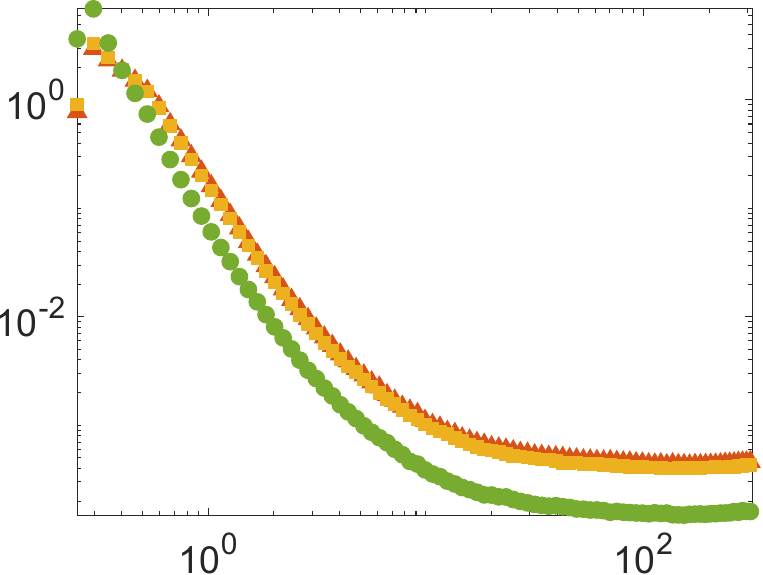}
                
                \put(90,60){$St = 9.3$}
                \put(-10,60){\rotatebox{90}{$\varrho$}}
                \put(90,-5){$z$}
		\end{overpic}}
        \subfloat[]
		{\begin{overpic}
				[trim = -15mm 0mm 0mm 0mm,scale=0.42,clip,tics=20]{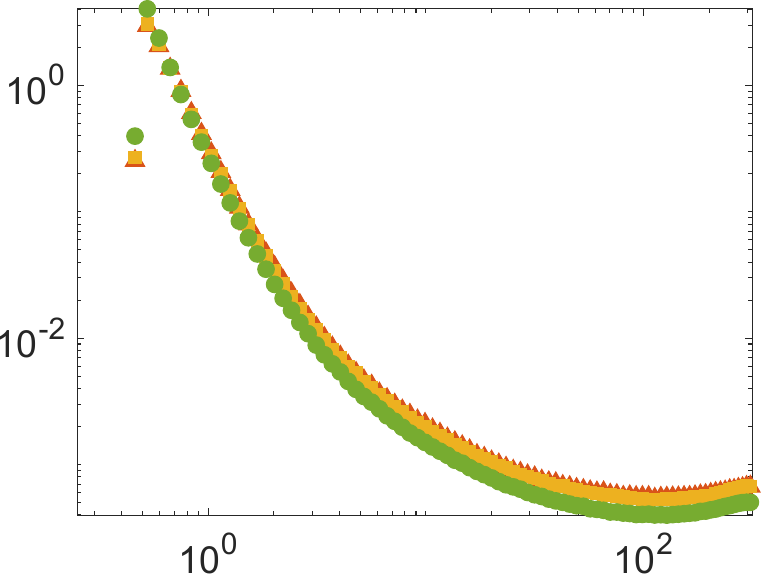}
                
                \put(110,60){$St = 46.5$}
                \put(10,60){\rotatebox{90}{$\varrho$}}
                \put(90,-5){$z$}
		\end{overpic}}
        \caption{Particle concentration $\varrho$ at different $Sv$ for (a) $St=0.93$, (b) $St=4.65$, (c) $St=9.3$, (d) $St=46.5$ in the HZF case.}
		\label{BC_Elas}
\end{figure}}

In Fig \ref{BC_Absp} the particle concentration profiles for HNF cases are shown for different $St, Sv$ values. When raising $Sv$ from $Sv = 10^{-4}$ to $Sv = 10^{-3}$, both of which correspond to the weak-settling regime, the near-wall concentration $\varrho$ is significantly affected by the change in $Sv$, and decreases as $Sv$ is increased. This is consistent with the analysis in Bragg \textit{et al.} \cite{Bragg2021} that demonstrated that the near-wall transport of inertial particles can be strongly affected by settling even when $Sv$ is very small. In the near wall region for the HNF cases, settling enables the particles to move almost ballistically through the near-wall region, leading to a flattened concentration profile  in this region.  

In Fig \ref{BC_Elas} the particle concentration profiles for HZF cases are shown. For this configuration we find that $Sv$ plays an even more significant role when $Sv\ll1$ compared to the HNF cases just considered in Fig. \ref{BC_Absp}. Indeed, we find that we had to use values of $Sv$ that are in general an order of magnitude smaller than those for the HNF cases in Fig \ref{BC_Absp} since otherwise the particle statistics do not reach a statistically steady state (instead they all slowly accumulate at the wall). In Fig. \ref{BC_Elas}(a) only $\varrho$ at $Sv = 0$ and $10^{-5}$ are shown, because even at $Sv = 10^{-4}$, settling is significant enough to prevent the statistics from reaching a steady state. This issue is highlighted in Fig. \ref{0d93_Converg} which shows $\varrho_{max}$ as a function of time for $St=0.93$ and (a) $Sv=10^{-5}$ and (b) $Sv=10^{-4}$. Whereas the case with $Sv = 10^{-5}$ reaches a steady state after approximately $t= 1\times 10^{5}\tau_*$, for the case with $Sv = 10^{-4}$, $\varrho_{max}$ continues to slowly increase with time and does not reach a steady state. This is associated with all of the particles slowly drifting towards the wall and remaining trapped there, which would appear to be the fate of them all for $t\to\infty$.

{\vspace{0mm}\begin{figure}
		\centering
        \subfloat[]
		{\begin{overpic}
				[trim = 0mm -5mm -10mm 0mm,scale=0.42,clip,tics=20]{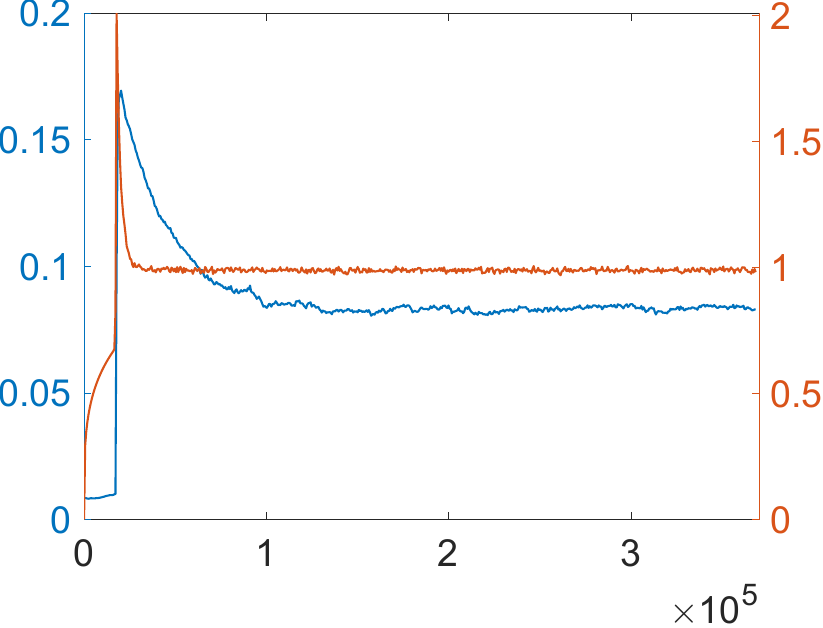}

                \put(80,0){$t/\tau_*$}
                \put(-10,70){\rotatebox{90}{$\varrho_{max}$}}
                \put(165,70){\rotatebox{90}{$u_\tau$}}
                \put(90,37){$Sv = 10^{-5}$}
                \put(90,60){$\varrho_{max}$}
                \put(90,85){$u_\tau$}
									
		\end{overpic}}
        \hspace{5mm}
        \subfloat[]
        {\begin{overpic}
				[trim = -10mm -5mm -10mm 0mm,scale=0.42,clip,tics=20]{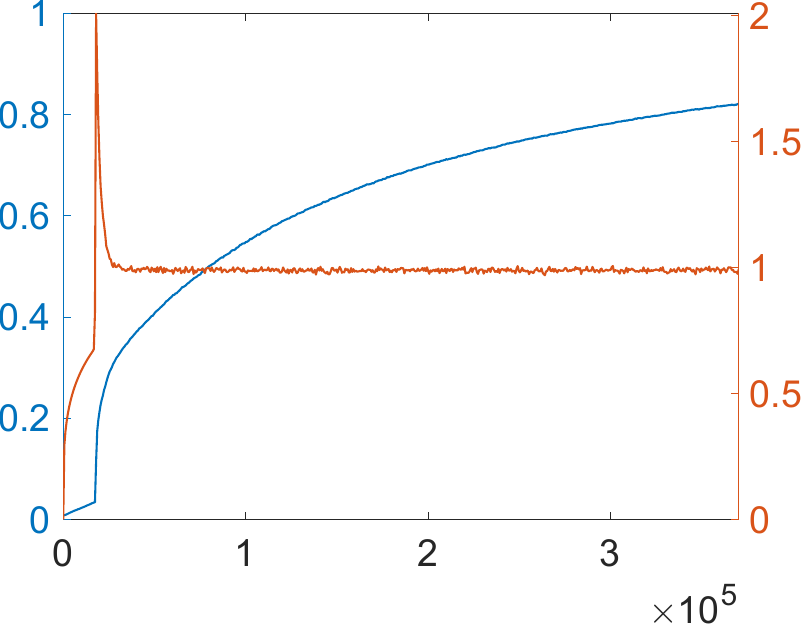}

                \put(85,0){$t/\tau_*$}
                \put(0,70){\rotatebox{90}{$\varrho_{max}$}}
                \put(170,70){\rotatebox{90}{$u_\tau$}}
                \put(90,37){$Sv = 10^{-4}$}
                \put(90,110){$\varrho_{max}$}
                \put(90,70){$u_\tau$}
		\end{overpic}}
        \caption{Maximum particle concentration $\varrho_{max}$ as a function of time $t$ for $St = 0.93$ in a HZF case (a) $Sv = 10^{-5}$, (b) $Sv = 10^{-4}$.  }
		\label{0d93_Converg}
\end{figure}
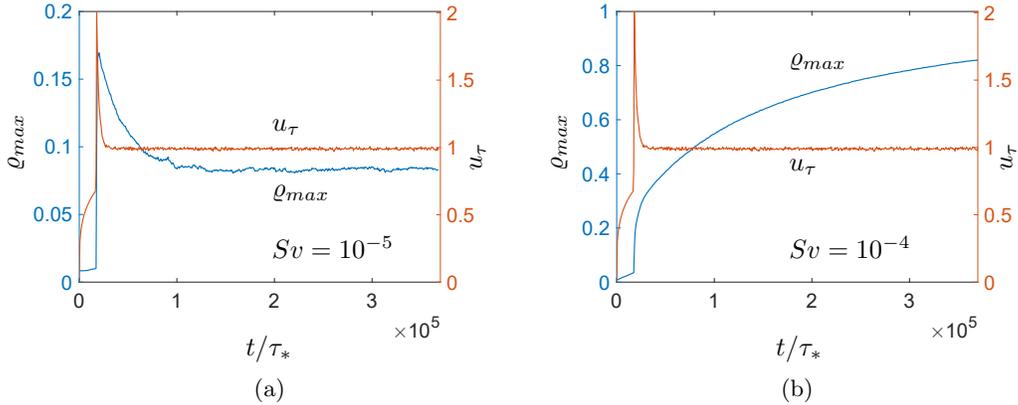}

In Figs. \ref{BC_Elas} (b), (c) \& (d) three small $Sv$ numbers ($Sv \in [ 0, 10^{-5}, 10^{-4} ]$) are tested, for which it is shown that the concentration is significantly affected by $Sv$ when going from $10^{-5}$ to $10^{-4}$. The sensitivity to $Sv$ is relatively weak for the case with $St=46.5$, however. This reflects the fact that when the particle inertia is sufficiently large, inertial effects dominate in the very near wall region, and settling would only play a large role in that region for larger $Sv$. The difference between the results for $Sv=0$ and $10^{-5}$ is in general negligible, however. This is reassuring since the particle equation of motion is regularly perturbed with respect to $Sv$ and therefore although the particle motion might be sensitive to $Sv$ even when $Sv$ is very small, this sensitivity must vanish in the limit $Sv\to 0$. The results suggest that the particle motion is extremely sensitive to $Sv$, however, and that values as low as $10^{-5}$ are required for the effect of settling to be truly negligible. We again stress that this is not only an issue for the near-wall region, but for the particle concentration throughout the entire flow, because the integral constraint on $\varrho$ means that any effects of $Sv$ on $\varrho$ in the near-wall region are also indirectly felt further away from the wall. This is confirmed in Figs. \ref{BC_Elas} (b), (c) \& (d), with the modification of $\varrho$ in the near-wall region due to settling also causing $\varrho$ to be modified further away from the wall so that the integral constraint $\int^{H-d_p/2}_{d_p/2}\varrho(z)\,dz=1$ is preserved.

Whereas for $\Phi<0$, increasing $Sv$ led to a decrease in the near-wall particle concentration, for $\Phi=0$ the opposite occurs. This is in agreement with the observations made earlier regarding equation \eqref{rhosln_absp}. Clearly, according to the comparison between the cases with $\Phi=0$ and $\Phi<0$, the effect of $Sv$ on $\varrho$ depends strongly on the particle boundary conditions which determine $\Phi$, with the effect being different not only quantitatively, but also qualitatively. This agrees with the discussion based on the theoretical result in equation \eqref{rhosln_elas}.

It is also worth noting that the results in Fig.  \ref{BC_Absp} \& Fig. \ref{BC_Elas} show that for some cases, the highest particle concentration occurs at a small distance from the wall, including for the cases with $Sv=0$. This has also been observed in previous DNS of non-settling inertial particles in turbulent channel flow \cite{Marchioli2008}. It is possible that this effect is physical, and is caused by the inertial particle velocities being too large relative to the fluid velocities very close to the wall for the particles to accumulate there, and so they accumulate at a small distance from the wall where the drag forces are more effective in capturing them. On the other hand, it could be due to insufficient statistical convergence very close to the wall and/or a finer grid being needed close to the wall. This may be the case since the asymptotic results in \citet{sikovsky14} suggest that for $z\to 0$ and $Sv=0$ the concentration $\varrho$ should grow as a power law.

\subsection{Explicit and implicit effect of settling on particle concentrations}
{\vspace{0mm}\begin{figure}
		\centering
		\subfloat[]
		{\begin{overpic}
				[trim = 0mm 0mm -15mm 0mm,scale=0.42,clip,tics=20]{figure4a.pdf}

                \put(25,140){\textbf{\large{Horizontal Channel}}}
                
                \put(100,60){$St = 0.93$}
                \put(-10,60){\rotatebox{90}{$\varrho$}}
                \put(80,-5){$z$}
									
		\end{overpic}}
		\subfloat[]
		{\begin{overpic}
				[trim = -15mm 0mm 0mm 0mm,scale=0.42,clip,tics=20]{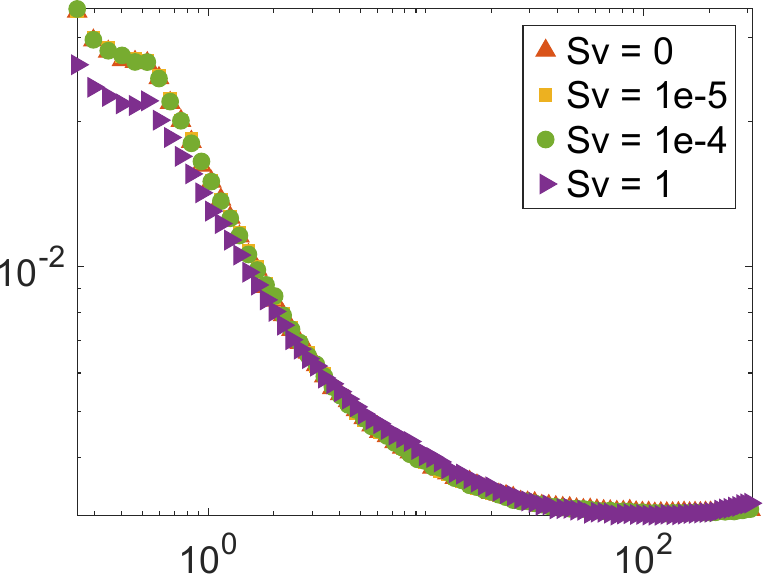}
                
                \put(55,140){\textbf{\large{Vertical Channel}}}
                
                \put(110,60){$St = 0.93$}
                \put(10,60){\rotatebox{90}{$\varrho$}}
                \put(100,-5){$z$}
                
		\end{overpic}}
        \\
        \subfloat[]
		{\begin{overpic}
				[trim = 0mm 0mm -15mm 0mm,scale=0.42,clip,tics=20]{figure4b.pdf}
                
                \put(100,60){$St = 4.65$}
                \put(-10,60){\rotatebox{90}{$\varrho$}}
                \put(80,-5){$z$}
		\end{overpic}}
        \subfloat[]
		{\begin{overpic}
				[trim = -15mm 0mm 0mm 0mm,scale=0.42,clip,tics=20]{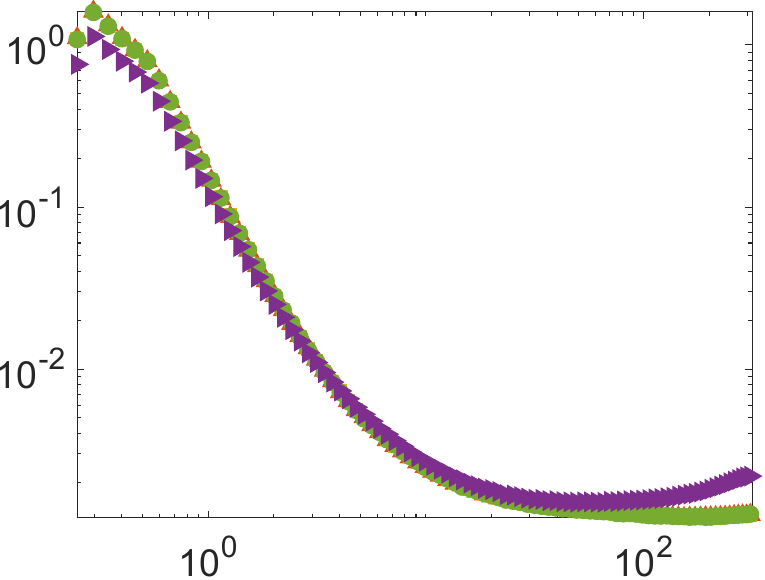}
                
                \put(100,60){$St = 4.65$}
                \put(10,60){\rotatebox{90}{$\varrho$}}
                \put(100,-5){$z$}
		\end{overpic}}
        \\
		\subfloat[]
		{\begin{overpic}
				[trim = 0mm 0mm -15mm 0mm,scale=0.42,clip,tics=20]{figure4c.pdf}
                
                \put(100,60){$St = 9.3$}
                \put(-10,60){\rotatebox{90}{$\varrho$}}
                \put(80,-5){$z$}
		\end{overpic}}
		\subfloat[]
		{\begin{overpic}
				[trim = -15mm 0mm 0mm 0mm,scale=0.42,clip,tics=20]{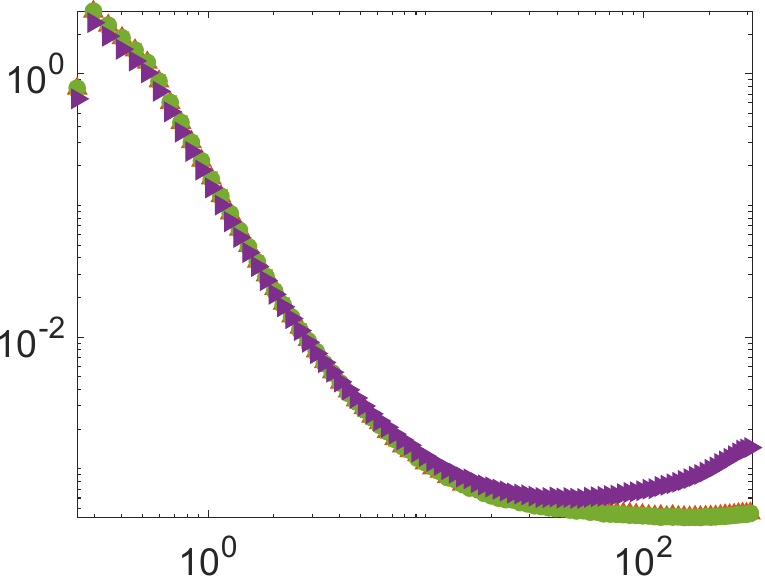}
                
                \put(110,60){$St = 9.3$}
                \put(10,60){\rotatebox{90}{$\varrho$}}
                \put(100,-5){$z$}
		\end{overpic}}
        \\
        \subfloat[]
		{\begin{overpic}
				[trim = 0mm 0mm -15mm 0mm,scale=0.42,clip,tics=20]{figure4d.pdf}
                
                \put(100,60){$St = 46.5$}
                \put(-10,60){\rotatebox{90}{$\varrho$}}
                \put(80,-5){$z$}
		\end{overpic}}
        \subfloat[]
		{\begin{overpic}
				[trim = -15mm 0mm 0mm 0mm,scale=0.42,clip,tics=20]{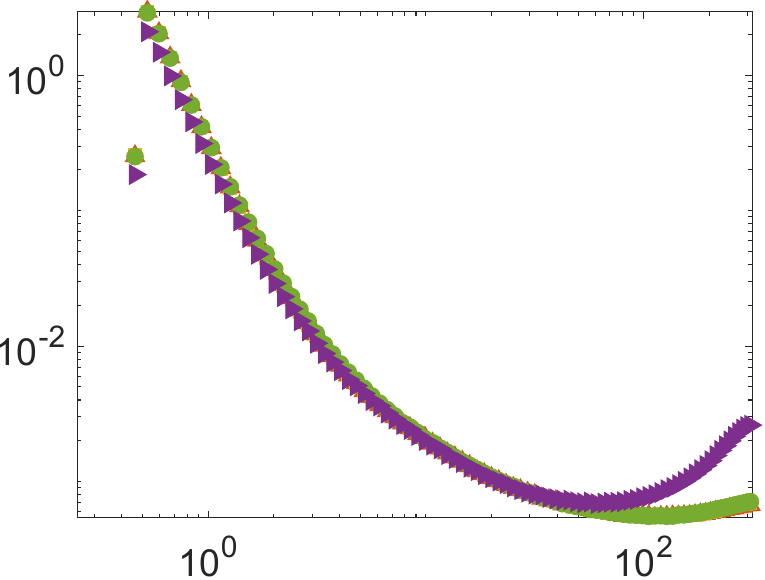}
                
                \put(100,60){$St = 46.5$}
                \put(10,60){\rotatebox{90}{$\varrho$}}
                \put(100,-5){$z$}
		\end{overpic}}
        \caption{DNS results of $\varrho$ at at (a,b) $St = 0.93$, (c,d) $St = 4.65$, (e,f) $St = 9.3$ and (g,h) $St = 46.5$ in log-log scale, (a),(c),(e),(g) for HZF and (b),(d),(f),(h) VZF cases.}
		\label{horivert_rho}
\end{figure}
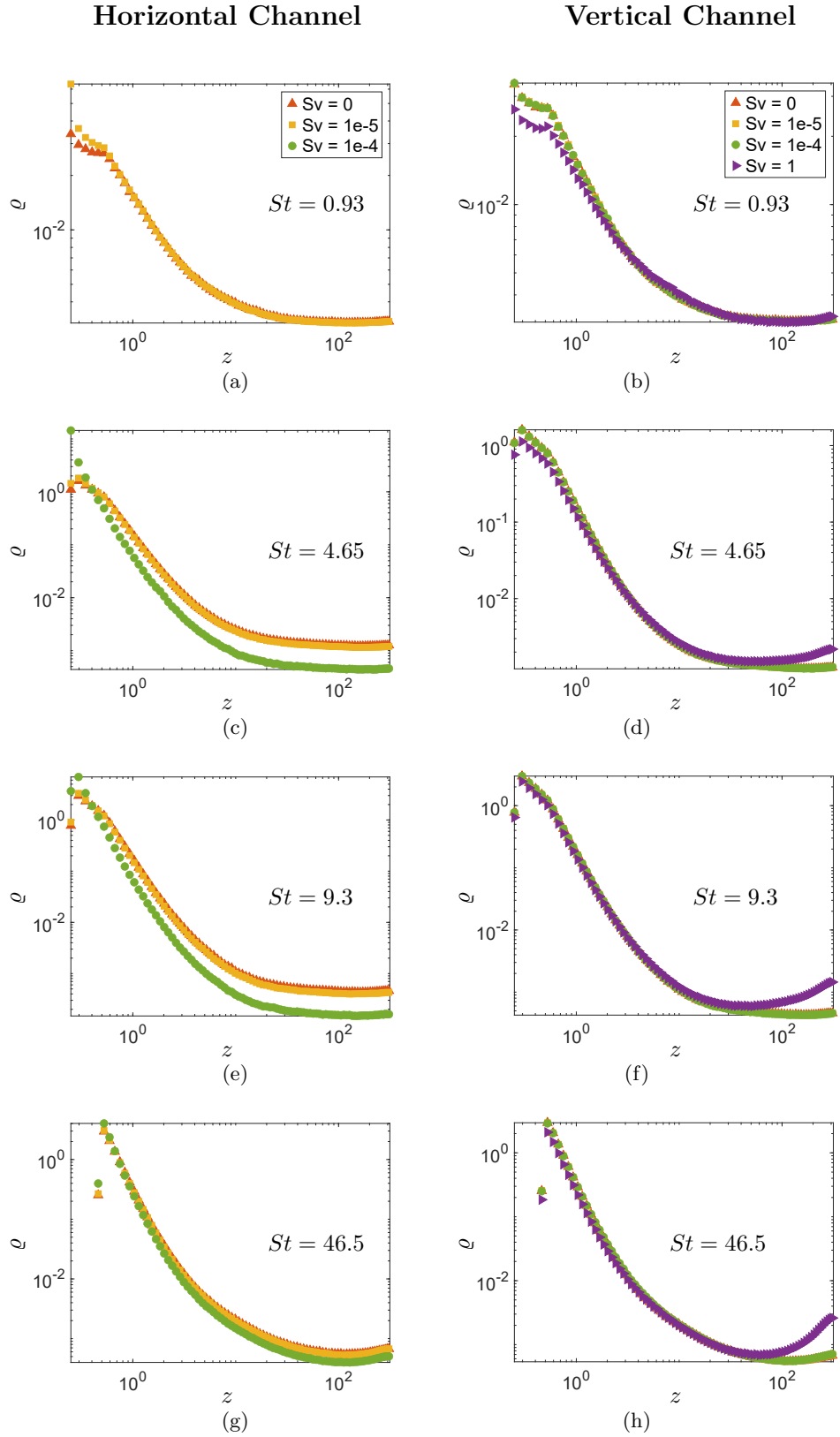}

As discussed in Section II, settling can impact $\varrho$ either explicitly or implicitly, depending on the orientation of the wall relative to gravity. In order to assess the relative importance of these effects, we compare the results from both the HZF (where both the explicit and implicit effects of settling play a role) and VZF (where only the implicit effect of settling plays a role) cases. The results are shown in Fig. \ref{horivert_rho}, and while the HZF results were already shown in Fig. \ref{BC_Elas}, they are included here again for direct comparison with the VZF cases. In contrast with the HZF cases where $\varrho$ is very sensitive to settling even when $Sv$ is small, for the same $Sv$ numbers the effect of settling on $\varrho$ is negligible for the VZF cases. This suggests that in the small $Sv$ regime the implicit effect of settling is negligible, and that the explicit effect is likely the dominant cause of the sensitivity of $\varrho$ to $Sv$ in both the HZF and HNF cases. This is consistent with the analysis in \cite{Bragg2021} which predicts that it is the explicit effect of settling that will cause the particle motion in horizontal flows to be sensitive to $Sv$ even when $Sv$ is small (but finite). 

The VZF cases in Fig. \ref{horivert_rho} also include results for $Sv=1$ in order to assess how large $Sv$ has to be in vertical channels before the implicit effect of settling starts to impact $\varrho$. The results show noticeable differences between the $Sv=0$ and $Sv=1$ cases, with the greatest differences occurring away from the wall and closer to the channel core. A practical implication of these results is that while the neglect of settling in models and simulations of particle transport in horizontal wall-bounded flows is not justified even when $Sv\ll1$ (unless  $Sv$ is extremely small, e.g. $\leq O(10^{-5})$), it is reasonable for vertical wall-bounded flows unless $Sv\geq O(1)$. 

\subsection{Mechanisms Governing Settling in a Vertical Channel}

{\vspace{0mm}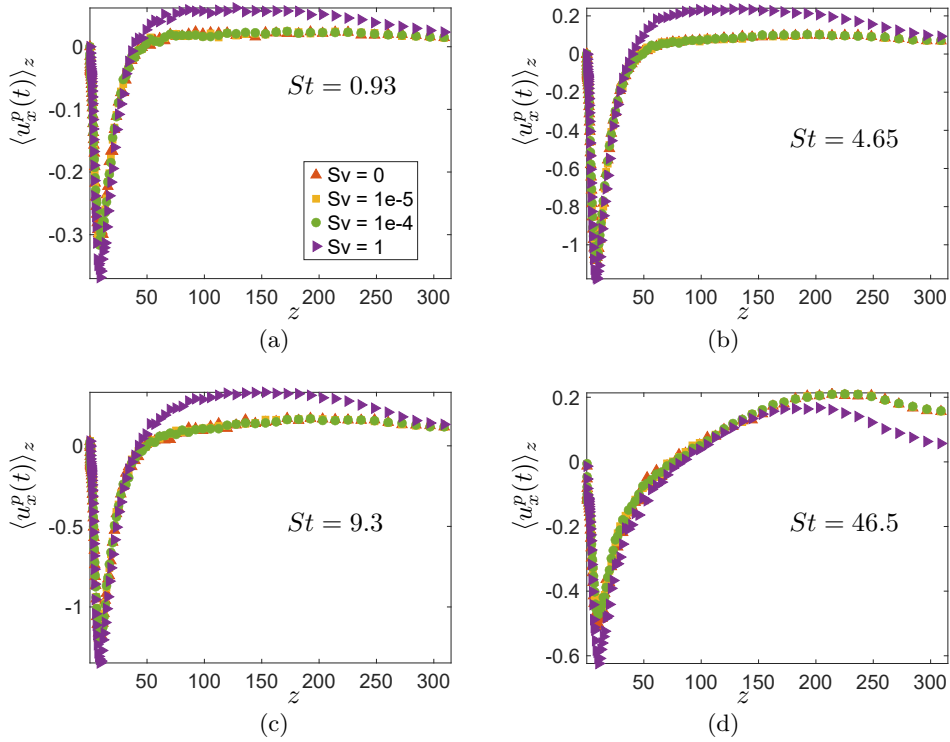
\begin{figure}
		\centering
		\subfloat[]
		{\begin{overpic}
				[trim = 0mm 0mm -15mm 0mm,scale=0.42,clip,tics=20]{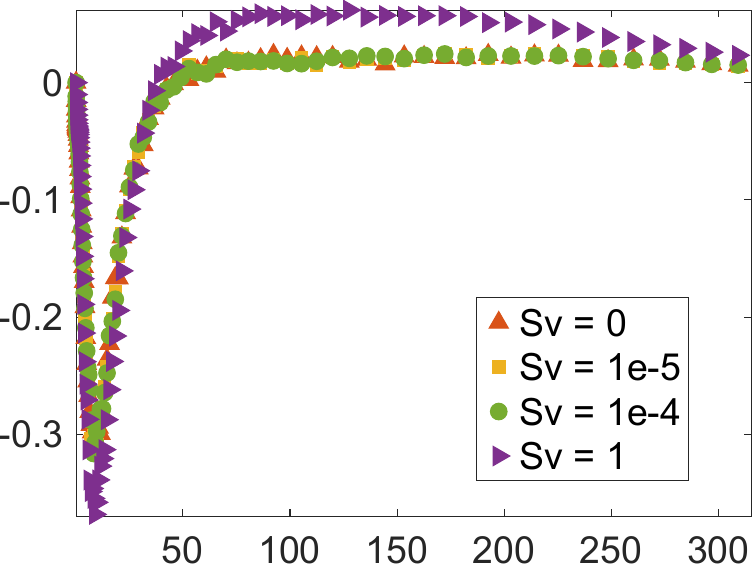}

                %\put(25,140){\textbf{\large{Elastic BC}}}
                
                \put(90,80){$St = 0.93$}
                % \put(40,80){$S_{xz}$}

                \put(-15,60){\rotatebox{90}{$\langle u_x^p(t)\rangle_z$}}
                \put(90,-5){$z$}
									
		\end{overpic}}
		\subfloat[]
		{\begin{overpic}
				[trim = -15mm 0mm 0mm 0mm,scale=0.42,clip,tics=20]{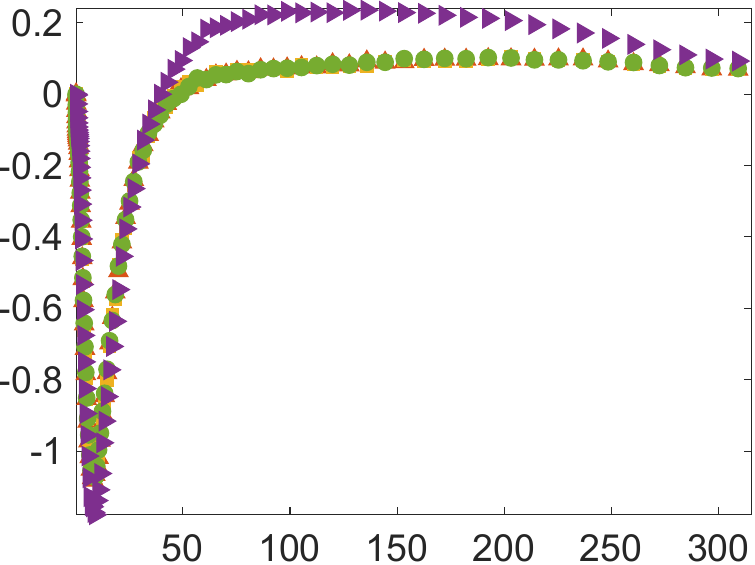}
                
                %\put(35,140){\textbf{\large{Absorbing BC}}}
                
                \put(110,60){$St = 4.65$}

                \put(5,60){\rotatebox{90}{$\langle u_x^p(t)\rangle_z$}}
                \put(90,-5){$z$}
                
		\end{overpic}}
        \\
        \subfloat[]
		{\begin{overpic}
				[trim = 0mm 0mm -15mm 0mm,scale=0.42,clip,tics=20]{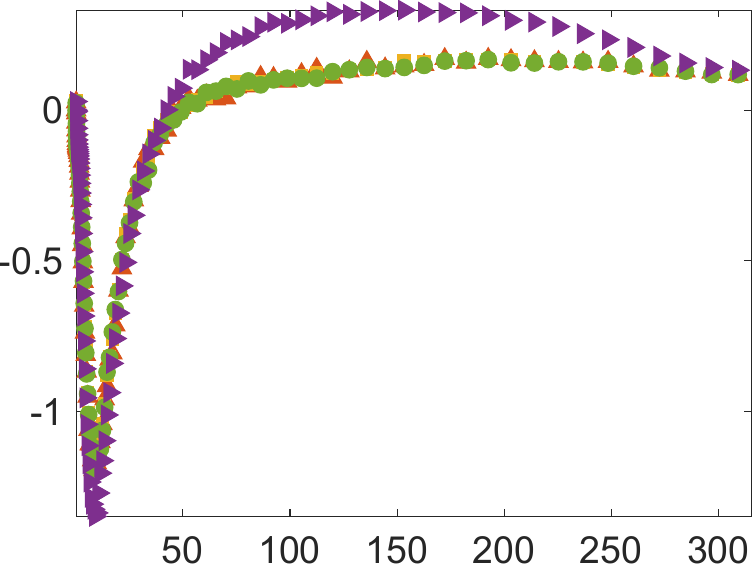}
                
                \put(90,60){$St = 9.3$}
                
                \put(-15,60){\rotatebox{90}{$\langle u_x^p(t)\rangle_z$}}
                \put(90,-5){$z$}
		\end{overpic}}
        \subfloat[]
		{\begin{overpic}
				[trim = -15mm 0mm 0mm 0mm,scale=0.42,clip,tics=20]{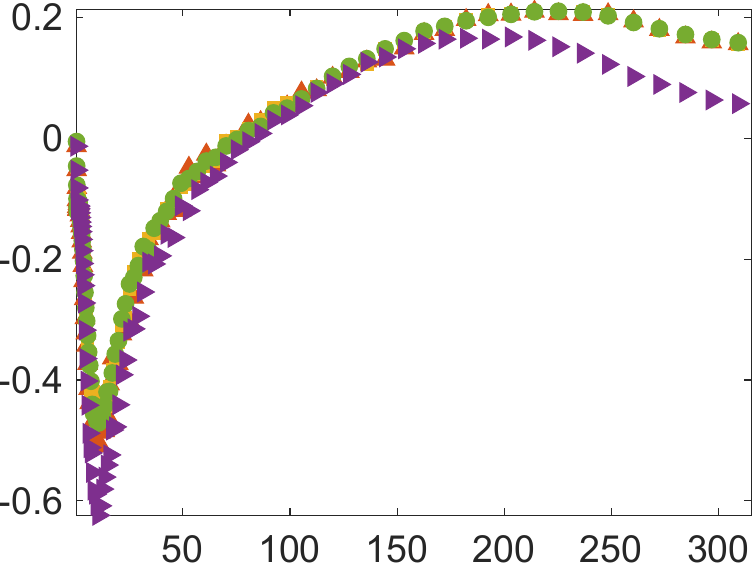}
                
                \put(110,60){$St = 46.5$}

                \put(5,60){\rotatebox{90}{$\langle u_x^p(t)\rangle_z$}}
                \put(90,-5){$z$}
		\end{overpic}}
        \caption{Plots of the fluctuating streamwise fluid velocity averaged at the particle positions $\langle u_x^p(t)\rangle_z$ at (a) $St = 0.93$, (b) $St = 4.65$, (c) $St = 9.3$, (d) $St = 46.5$}
		\label{Vert_R4x}
\end{figure}}

In \cite{bragg21} we explored in detail the mechanisms governing the settling velocity of inertial particles in HNF cases, where the settling is in the wall-normal direction. We now consider this for our VZF cases where the settling is in the streamwise direction. For the VZF cases, the exact steady-state equation (corresponding to the equation of motion in equation \ref{parteq}) for the mean particle streamwise velocity can be derived from the phase-space PDF equation and is \cite{bragg21}
\begin{align}
\langle w_x^p(t)\rangle_z=-\frac{St}{\varrho}S_{xz}\nabla_z\varrho -St\nabla_z S_{xz} +\langle u_x^p(t)\rangle_z + U_x+Sv,  \label{wx_eq}
\end{align}
where $w_x^p(t)$ is the streamwise component of the particle velocity, $S_{xz}\equiv\langle[w_x^p(t)-\langle w_x^p(t)\rangle_z][w^p(t)-\langle w^p(t)\rangle_z]\rangle_z$ is the covariance of the streamwise and wall-normal particle velocity, $u_x^p(t)$ is the fluctuating streamwise fluid velocity at the particle position, and $U_x$ is the mean streamwise fluid velocity evaluated at the particle position. Far from the wall where gradients in the turbulent statistics are weak, the equation reduces to $\langle w_x^p(t)\rangle_z\sim\langle u_x^p(t)\rangle_z + U_x+Sv$ (unless $St$ is very large, i.e. large enough for the particles to move ballistically through the channel core) and in this region we would expect $\langle u_x^p(t)\rangle_z>0$ due to the preferential sweeping mechanism. The results in Fig. \ref{Vert_R4x} confirm this expectation. Close to the wall, however, the results show that $\langle u_x^p(t)\rangle_z<0$. For $Sv=0$ this is known to be associated with the accumulation of the inertial particles in low-speed streaks in the boundary layer \cite{RASHIDI1990935,eaton94,marchioli02}, and our results show that this continues to be the case even for $Sv=O(1)$. The implication of this is that for particles settling near a vertical boundary, the preferential sampling of low speed streaks acts to suppress their settling speed, in contrast to their settling far from boundaries where preferential sampling of downward moving strain-dominated regions of the flow enhances their settling speed through the preferential sweeping mechanism.

{\vspace{0mm}\begin{figure}
		\centering
		\subfloat[]
		{\begin{overpic}
				[trim = 0mm 0mm -15mm 0mm,scale=0.42,clip,tics=20]{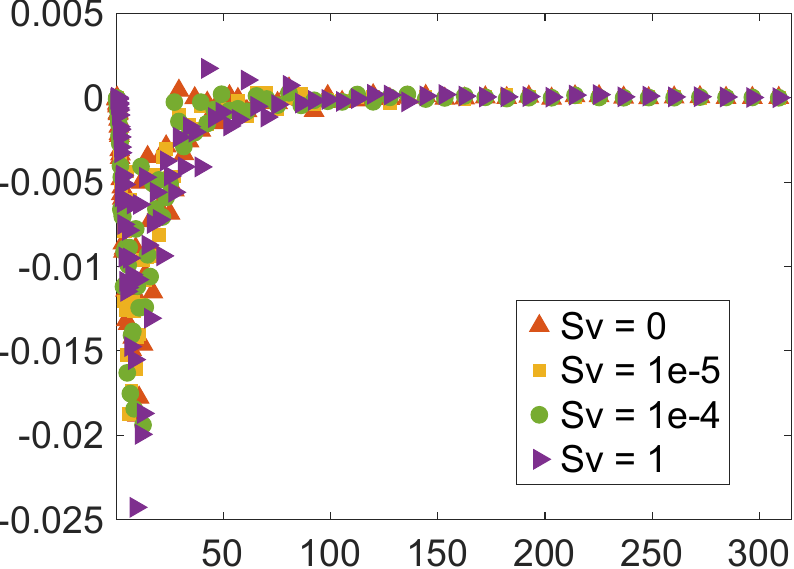}

                %\put(25,140){\textbf{\large{Elastic BC}}}
                
                \put(90,80){$St = 0.93$}
                % \put(40,80){$S_{xz}$}
                
                \put(-15,30){\rotatebox{90}{$-(St/\varrho) S_{xz}\nabla_z \varrho$}}
                \put(90,-5){$z$}
									
		\end{overpic}}
		\subfloat[]
		{\begin{overpic}
				[trim = -15mm 0mm 0mm 0mm,scale=0.42,clip,tics=20]{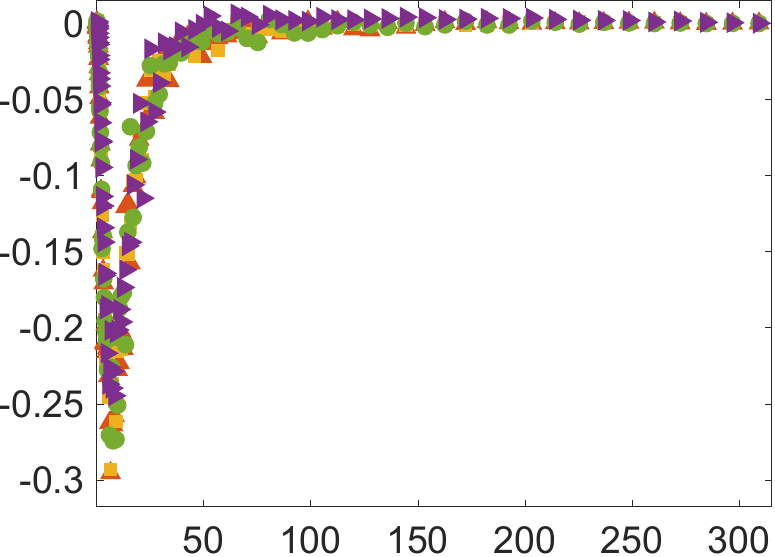}
                
                %\put(35,140){\textbf{\large{Absorbing BC}}}
                
                \put(110,60){$St = 4.65$}
                
                \put(5,30){\rotatebox{90}{$-(St/\varrho)S_{xz}\nabla_z \varrho$}}
                \put(90,-5){$z$}
                
		\end{overpic}}
        \\
        \subfloat[]
		{\begin{overpic}
				[trim = 0mm 0mm -15mm 0mm,scale=0.42,clip,tics=20]{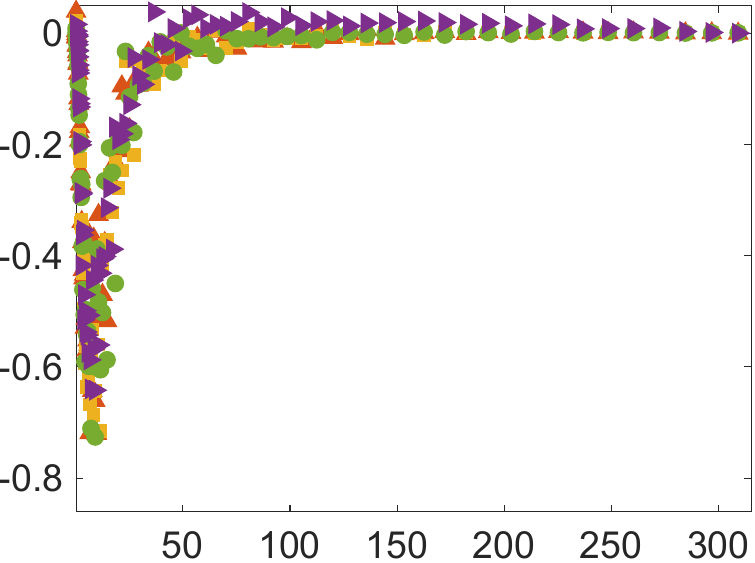}
                
                \put(90,60){$St = 9.3$}
                \put(-15,30){\rotatebox{90}{$-(St/\varrho) S_{xz}\nabla_z \varrho$}}
                \put(90,-5){$z$}
		\end{overpic}}
        \subfloat[]
		{\begin{overpic}
				[trim = -15mm 0mm 0mm 0mm,scale=0.42,clip,tics=20]{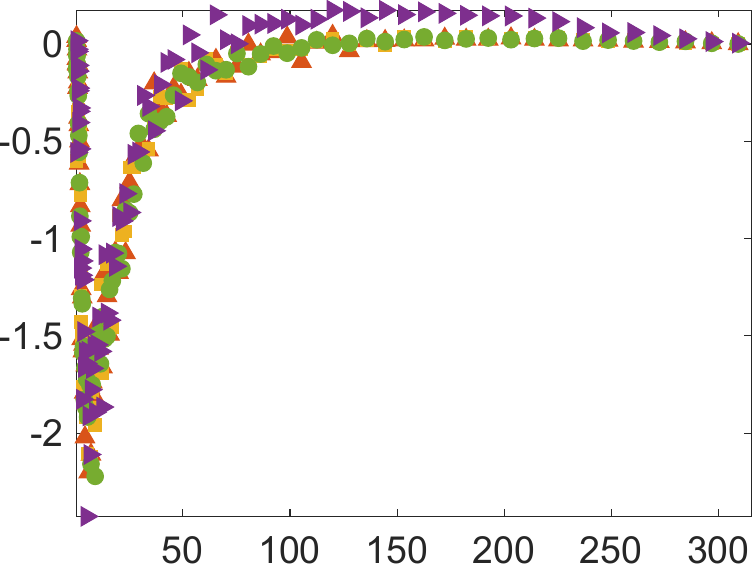}
                
                \put(110,60){$St = 46.5$}
                
                \put(5,30){\rotatebox{90}{$-(St/\varrho)S_{xz}\nabla_z \varrho$}}
                \put(90,-5){$z$}
		\end{overpic}}
        \caption{Plots of the diffusion term $-(St/\varrho) S_{xz}\nabla_z \varrho$ at (a) $St = 0.93$, (b) $St = 4.65$, (c) $St = 9.3$, (d) $St = 46.5$}
		\label{Vert_R2x}
\end{figure}}

{\vspace{0mm}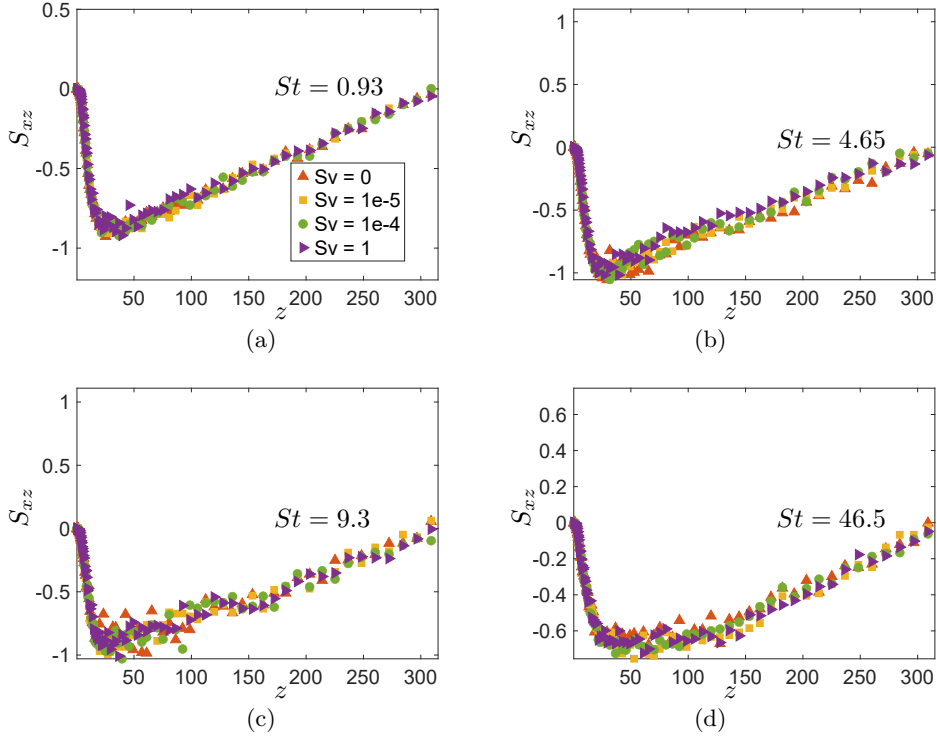
\begin{figure}
		\centering
		\subfloat[]
		{\begin{overpic}
				[trim = 0mm 0mm -15mm 0mm,scale=0.42,clip,tics=20]{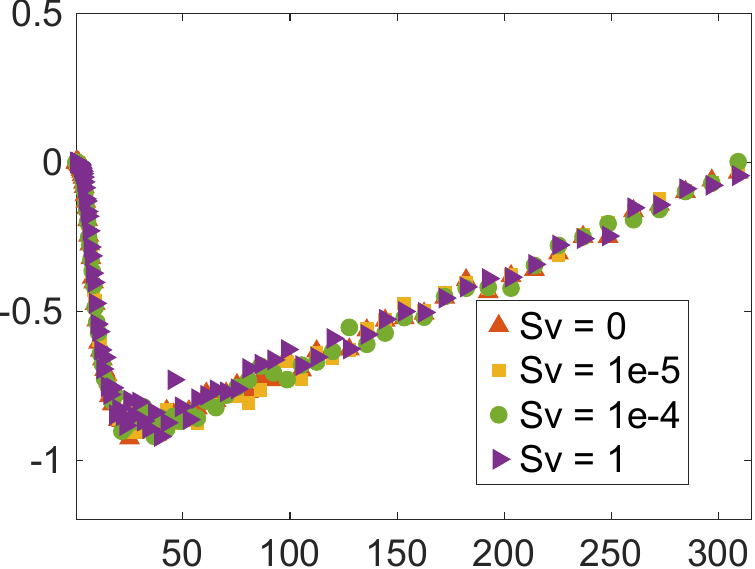}

                %\put(25,140){\textbf{\large{Elastic BC}}}
                
                \put(90,80){$St = 0.93$}
                %\put(40,80){$S_{xz}$}
                \put(-10,60){\rotatebox{90}{$S_{xz}$}}
                \put(90,-5){$z$}
									
		\end{overpic}}
		\subfloat[]
		{\begin{overpic}
				[trim = -15mm 0mm 0mm 0mm,scale=0.42,clip,tics=20]{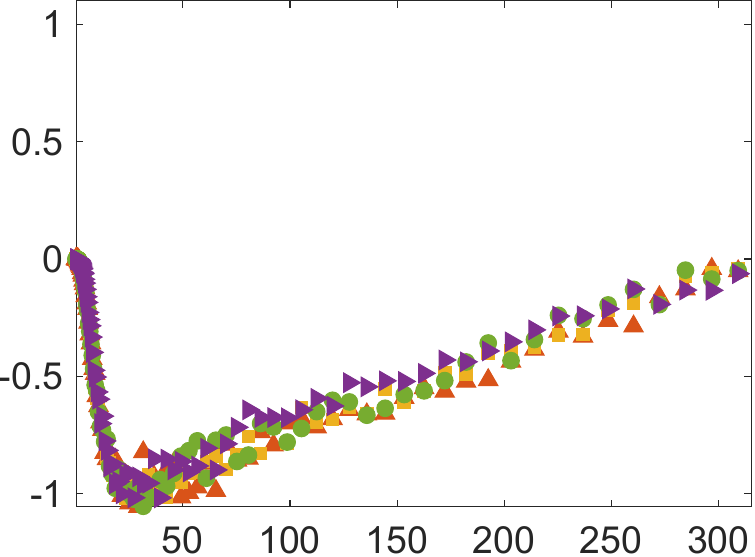}
                
                %\put(35,140){\textbf{\large{Absorbing BC}}}
                
                \put(110,60){$St = 4.65$}
                \put(10,60){\rotatebox{90}{$S_{xz}$}}
                \put(90,-5){$z$}
                
		\end{overpic}}
        \\
        \subfloat[]
		{\begin{overpic}
				[trim = 0mm 0mm -15mm 0mm,scale=0.42,clip,tics=20]{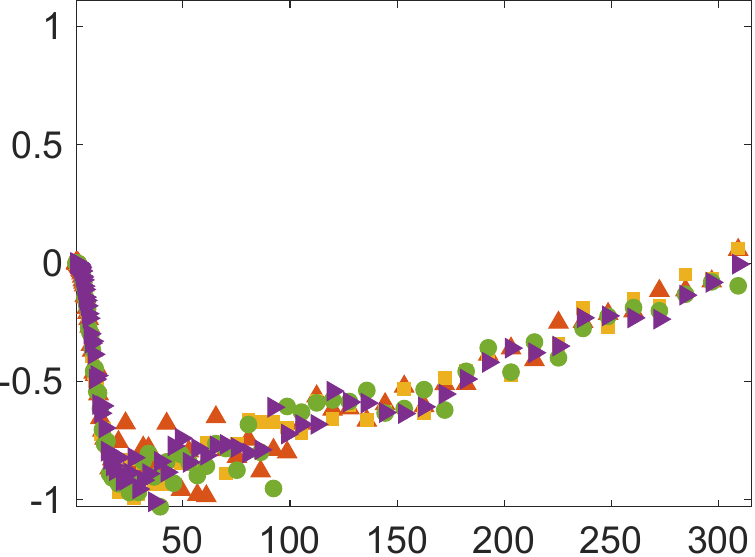}
                
                \put(90,60){$St = 9.3$}
                \put(-10,60){\rotatebox{90}{$S_{xz}$}}
                \put(90,-5){$z$}
		\end{overpic}}
        \subfloat[]
		{\begin{overpic}
				[trim = -15mm 0mm 0mm 0mm,scale=0.42,clip,tics=20]{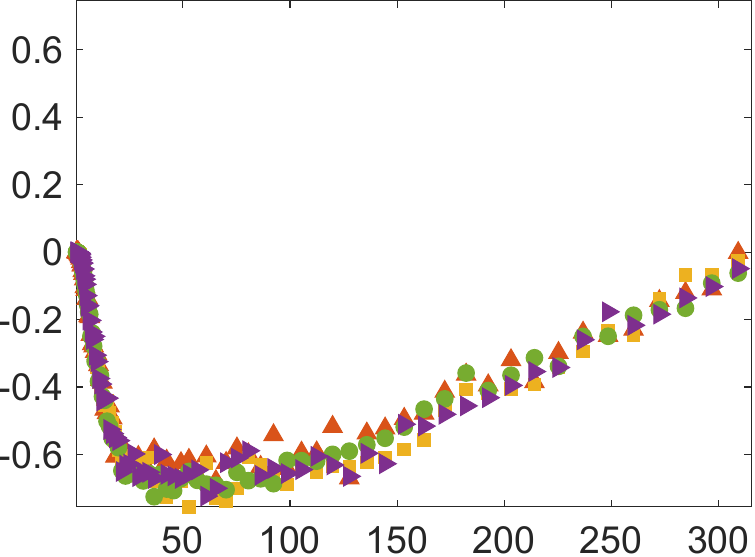}
                
                \put(110,60){$St = 46.5$}
                \put(10,60){\rotatebox{90}{$S_{xz}$}}
                \put(90,-5){$z$}
		\end{overpic}}
        \caption{Plots of the covariance $S_{xz}$ at (a) $St = 0.93$, (b) $St = 4.65$, (c) $St = 9.3$, (d) $St = 46.5$}
		\label{Vert_Sxz}
\end{figure}}

The results in Fig. \ref{Vert_R2x} show that close to the wall where the concentration gradients are strong, the diffusive contribution $-(St/\varrho)S_{xz}\nabla_z\varrho $ in \eqref{wx_eq} can be significant for larger $St$. This contribution is also negative, hindering the settling of the particles close to the wall together with $\langle u_x^p(t)\rangle_z$. The results in Fig. \ref{Vert_Sxz} imply that close to the wall the contribution $-St\nabla_z S_{xz} $ in \eqref{wx_eq} can also be significant for larger $St$ (due to noise in the data, the figure shows $S_{xz}$ rather than its gradient). This term is by contrast positive close to the wall, and therefore enhances the settling velocity of the particle. The physical interpretation of this term is similar to the standard turbophoretic drift velocity (see \cite{bragg21} for a detailed discussion), except that the standard turbophoretic drift involves exclusively wall-normal particle velocity fluctuations, whereas $-St\nabla_z S_{xz} $ involves wall-normal and streamwise particle fluctuating velocity components. In particular, it is associated with the particles transferring streamwise momentum in the wall-normal direction due to their ability to remember their interaction with the flow in regions where the streamwise fluctuations were on average different (due to turbulence inhomogeneity) than those at their current location, thereby transferring streamwise momentum to their current location.

{\vspace{0mm}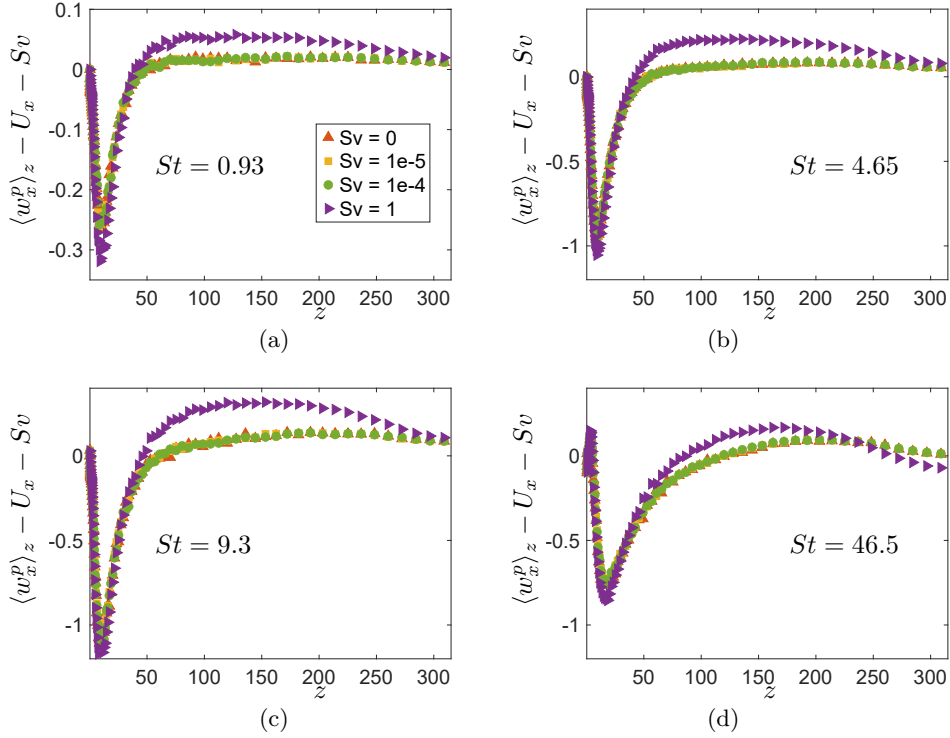
\begin{figure}
		\centering
		\subfloat[]
		{\begin{overpic}
				[trim = 0mm 0mm -15mm 0mm,scale=0.42,clip,tics=20]{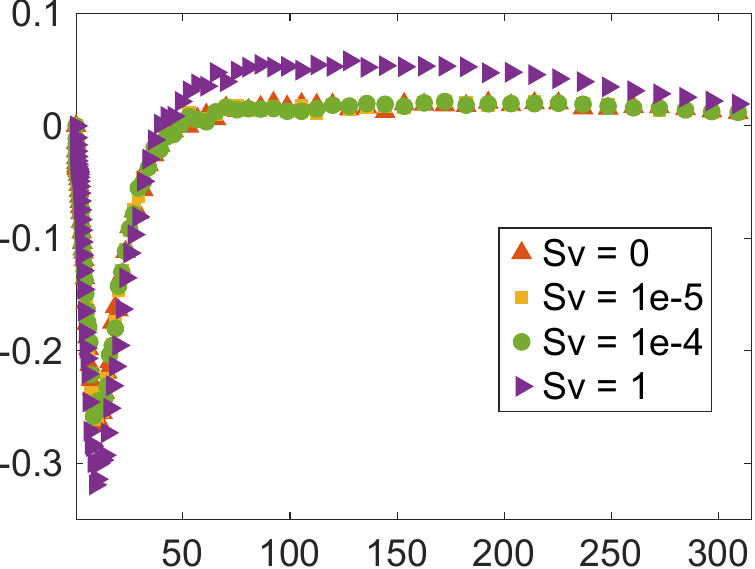}

                %\put(25,140){\textbf{\large{Elastic BC}}}
                
                \put(40,50){$St = 0.93$}
                %\put(50,75){$\langle w^p_{x}\rangle - \langle u\rangle - Sv$}
                
                \put(-15,30){\rotatebox{90}{$\langle w^p_{x}\rangle_z - U_x  - Sv$}}
                \put(100,-5){$z$}
									
		\end{overpic}}
		\subfloat[]
		{\begin{overpic}
				[trim = -15mm 0mm 0mm 0mm,scale=0.42,clip,tics=20]{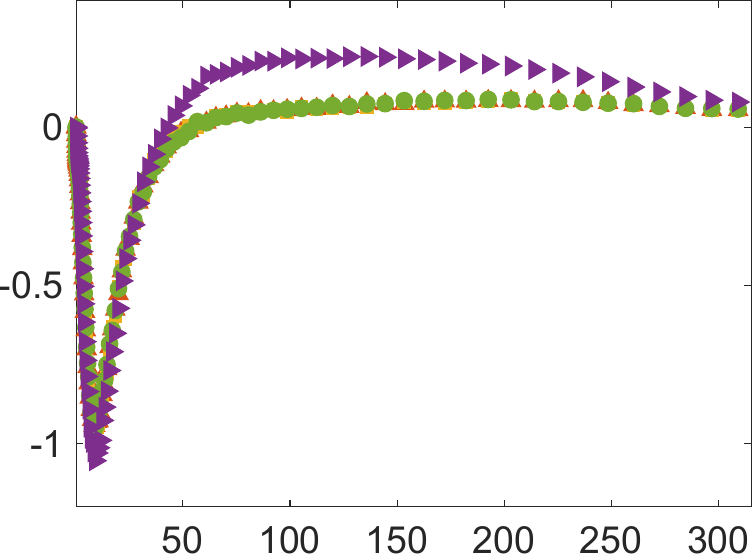}
                
                %\put(35,140){\textbf{\large{Absorbing BC}}}
                
                \put(110,50){$St = 4.65$}
                
                \put(5,30){\rotatebox{90}{$\langle w^p_{x}\rangle_z - U_x - Sv$}}
                \put(100,-5){$z$}
                
		\end{overpic}}
        \\
        \subfloat[]
		{\begin{overpic}
				[trim = 0mm 0mm -15mm 0mm,scale=0.42,clip,tics=20]{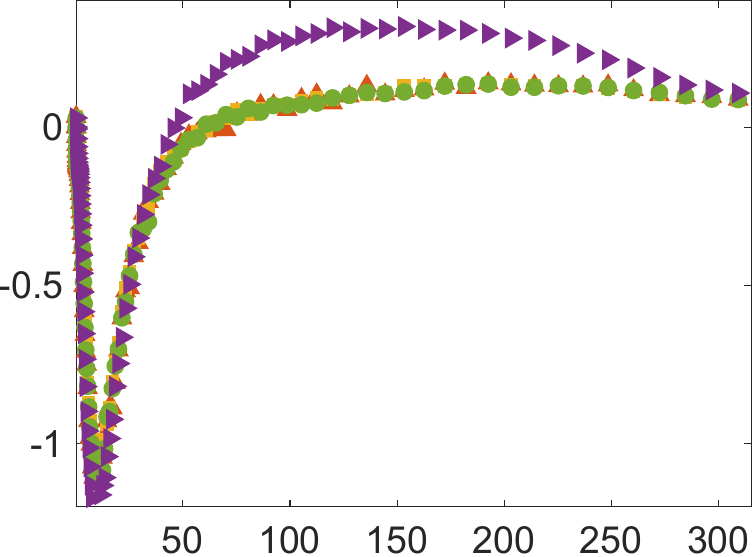}
                
                \put(40,50){$St = 9.3$}
                \put(-15,30){\rotatebox{90}{$\langle w^p_{x}\rangle_z -U_x - Sv$}}
                \put(100,-5){$z$}
		\end{overpic}}
        \subfloat[]
		{\begin{overpic}
				[trim = -15mm 0mm 0mm 0mm,scale=0.42,clip,tics=20]{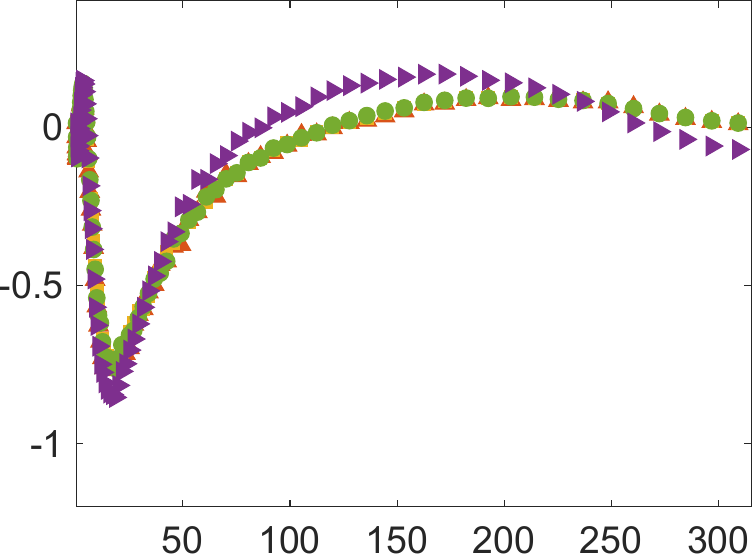}
                
                \put(110,50){$St = 46.5$}
                \put(5,30){\rotatebox{90}{$\langle w^p_{x}\rangle_z - U_x - Sv$}}
                \put(100,-5){$z$}
		\end{overpic}}
        \caption{Plots of the turbulence modification to the particle settling velocity $\langle w^p_{x}\rangle - U_x - Sv$ in a vertical channel at (a) $St = 0.93$, (b) $St = 4.65$, (c) $St = 9.3$, (d) $St = 46.5$}
		\label{Vert_wxbalance}
\end{figure}}

Finally, we consider the overall modification to the particle settling velocity in the vertical channel, $\langle w^p_{x}\rangle - U_x - Sv$, which corresponds to the first three terms on the rhs of \eqref{wx_eq}. As indicated earlier, further away from the wall where gradients in the particle statistics are weak, turbulence leads to an enhancement of the particle settling velocity through the preferential sweeping mechanism and $\langle w^p_{x}\rangle - U_x - Sv>0$, although the enhancement is negligible unless $Sv\geq O(1)$. Near the wall, however, turbulence significantly suppresses the settling of the inertial particles, and the results suggest that the preferential sampling of the low speed streaks through the sampling term $\langle u_x^p(t)\rangle_z $ makes an important contribution to this suppression process.

\section{Conclusions}

 In this work, we have considered how wall orientation and particle boundary conditions impact the transport of settling inertial particles in wall-bounded turbulent flows. This work was in part motivated by our recent study that found that, contrary to usual expectations, the concentrations of inertial particles could be strongly affected by settling even when the the settling parameter is $Sv\ll1$ \cite{Bragg2021}. That study was for particles in a horizontal channel flow with absorbing walls, and in this paper we have explored whether this same surprising behavior remains true for horizontal channel flows with elastic particle-wall collisions, and also whether it still occurs when the channel is vertical. When the channel is vertical, settling does not explicitly affect the wall-normal particle transport and concentration profiles, but it does affect them implicitly through the way that it influences how the particles interact with the turbulent flow field.

 These questions were examined both theoretically, by analyzing the exact phase-space PDF transport equations governing the particles, and numerically, using DNS. In each case, the particles were assumed to be point particles, governed by drag and gravitational forces. For the horizontal channels, it was found that the type of boundary condition strongly affects the sensitivity of the particle concentrations to $Sv$. In particular, the concentrations were far more sensitive to $Sv$ for elastic walls than absorbing walls, with the concentrations growing in the near wall region with increasing $Sv$ for the former, and decreasing for the latter. For the elastic walls, even when $Sv\ll1$, the effect of settling on the transport was so strong that the concentration did not even reach a steady state for some of the Stokes numbers. Furthermore, we found that values as small as $Sv=O(10^{-5})$ are required in order for the effect of settling on the particle concentrations to be truly negligible.

 We then compared the effect of settling on particle concentrations in horizontal and vertical channels (both having elastic particle-wall collisions). The results showed that in contrast to the horizontal case, the effect of settling is negligible when $Sv\ll1$ for the vertical case, and only becomes significant when $Sv\geq O(1)$. By implication, this means that the strong effect of settling in the horizontal cases even when $Sv\ll1$ must be due to the explicit effect of settling, with only a subleading contribution from the implicit effect. Unlike the horizontal channels, in the case of vertical channels, it is safe to neglect settling provided that $Sv\ll1$.

 Finally, we examined the mechanisms governing the settling velocity of the inertial particles in the case of a vertical channel, just as had been done for horizontal channels in \cite{bragg21}. Far from the wall, the turbulent contribution to the settling velocity is associated with the preferential sweeping mechanism which enhances the settling speed of the particles, just as it does for the horizontal channel. As the wall is approached, and the turbulence inhomogeneity becomes significant, the particles no longer preferentially sample downward moving strain dominated regions of the flow but instead preferentially sample regions where the fluctuating streamwise fluid velocity is negative. This has previously been observed for the $Sv=0$ case where is was associated with the preferential sampling of low speed streak structures in the boundary layer, and the same holds true for $Sv=O(1)$. The preferential sampling of such regions of the flow hinders the settling of the particles. Near the wall a diffusive mechanism associated with gradients in the particle concentration also acts to hinder the settling of the particles, while a mechanism analogous to the turbophoretic drift acts to enhance the settling velocity of the particles, and is associated with the transport of streamwise momentum in the wall-normal direction by the particles.

\section*{Acknowledgements}

Support from the Army Research Office (Award No. W911NF-22-2-0222) is gratefully acknowledged. We also thank Andrew P. Grace for help with using the DNS code.

 \bibliography{bib/Yanpaper2}

\end{document}